\documentclass[twocolumn,prb]{revtex4}
\usepackage[utf8]{inputenc}
\usepackage{amsmath,amsfonts,amssymb,graphicx,multirow,overpic,dsfont,bm,braket}
\usepackage{natbib}
\usepackage{xcolor}

\begin{document}

\title{Classification of subsystem symmetry-protected topological phases} 
\author{Trithep Devakul}
\affiliation{Department of Physics, Princeton University, Princeton, NJ 08540, USA}
\affiliation{%
Kavli Institute for Theoretical Physics, University of California, Santa Barbara, CA 93106, USA
}%
  \author{Dominic J. Williamson}
  \affiliation{Department of Physics, Yale University, New Haven, CT 06520-8120, USA}
  \author{Yizhi You}
  \affiliation{Princeton Center for Theoretical Science, Princeton University, 
NJ, 08544, USA}
\date{\today}
\date{\today}

\begin{abstract}
We consider symmetry-protected topological (SPT) phases in 2D protected by linear subsystem symmetries, i.e. those that act along rigid lines.
 There is a distinction between a ``strong'' subsystem SPT phase, and a ``weak'' one, which is composed of decoupled 1D SPTs with global symmetries.
 We propose a natural definition for \emph{strong equivalence} of such phases, in terms of a \emph{linearly-symmetric} local unitary transformation, under which a weak subsystem SPT is equivalent to the trivial phase.
 This leads to a number of distinct equivalence classes of strong subsystem SPTs, which we show are in one-to-one correspondence with elements of the group
 $\mathcal{C}[G_s] = \mathcal{H}^{2}[G_s^2,U(1)]/(\mathcal{H}^{2}[G_s,U(1)])^3$, where $G_s$ is the finite abelian onsite symmetry group.
 We also show that strong subsystem SPTs by our classification necessarily exhibit a spurious topological entanglement entropy on a cylinder.
\end{abstract}

\maketitle

\section{Introduction}
An important and long running goal of condensed matter physics is the enumeration and classification of all phases of matter.  
Much progress has been made toward this goal in the past decade, especially in the presence of symmetry.  
Even without long-range entanglement or Landau spontaneous symmetry breaking, there exists distinct gapped short-range entangled symmetric phases, known as symmetry-protected topological (SPT) phases~\cite{Chen2011-et,Chen2011-ss,Schuch2011-jx,Pollmann2010-fl,Senthil2015-tp,Chen2013-gq,Chen2012-tt}.

\begin{figure}[t]
    \centering
\includegraphics[width=0.30\textwidth]{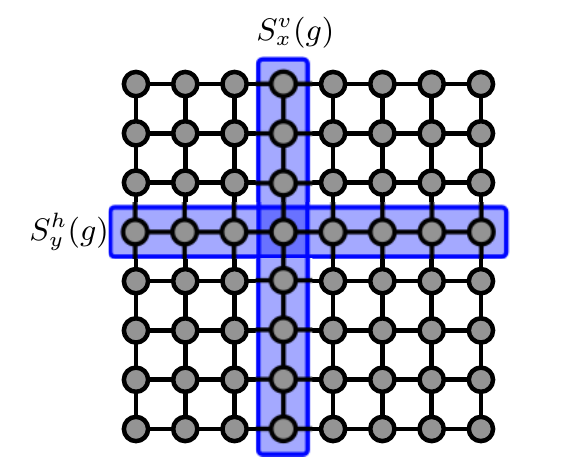}
    \caption{A depiction of the types of models we consider here.
    Each site transforms as a linear representation of some onsite symmetry group $G_s$.  
    For any $g\in G_s$, there is a vertical symmetry $S^v_x(g)$ on every column, and a horizontal symmetry $S^h_y(g)$ on every row, which apply the onsite representation of $g$ to every site in a column or row, respectively .
    We require a Hamiltonian to commute with all $S^v_x(g)$ and $S^h_y(g)$.
    }
    \label{fig:setting}
\end{figure}

The vast majority of SPT phases studied thus far possess \emph{global} symmetries.
These include well known examples such as the Haldane chain~\cite{Affleck1999-qv,Pollmann2012-lv} and topological insulators~\cite{Kane2005-eb}.
Recently, a new class of symmetries have been gaining attention in various different contexts:
these are symmetries which act non-trivially on only a vanishing fraction of the system.
These include higher form symmetries~\cite{Gaiotto2015-cj}, in which symmetries act on lower-dimensional \emph{deformable} manifolds, which have recently found application in quantum memories and error correction~\cite{Yoshida2016-zo,Kubica2018-dp,Roberts2017-uj,Roberts2018-xj}.
Of interest to us here are instead \emph{subsystem symmetries}, those which act on \emph{rigid} subsystems that cannot be deformed.
Such symmetries have historically also been referred to as intermediate or gauge-like symmetries, and have a variety of interesting properties~\cite{Batista2005-yr,Nussinov2009-ld,Nussinov2009-sw}.
A classic example is that of symmetries that act on rigid lines of the square lattice (see Fig.~\ref{fig:setting}), 
which arise in models of spin and orbital degrees of freedom, such as in the Kugel-Khomskii model~\cite{Kugel1973-bf,Van_den_Brink2004-og}, from Jahn-Teller effects~\cite{Van_den_Brink2004-og}, or in orbital compass models~\cite{Xu2004-oj}, which is dual to the Xu-Moore model of $p\pm ip$ superconducting arrays~\cite{Xu2005-df}.
More generally, subsystem symmetries include those that act on two-dimensional planes of a three dimensional model, such as in the plaquette Ising model~\cite{Johnston2012-rt,Vijay2016-dr, PhysRevB.81.184303}, or even on \emph{fractal} subsystems~\cite{Devakul2018-ru,Yoshida2013-of,Williamson2016-lv,Newman1999-fq,doi:10.1080/14786435.2011.609152}.

Such symmetries have witnessed a resurgence of interest owing to the discovery of \emph{fracton topological order}~\cite{Vijay2016-dr,Chamon2005-fc,Haah2011-ny,Bravyi2011-fl,Yoshida2013-of,Vijay2015-jj},
 a novel type of topological order characterized by the presence of a subextensive topological ground state degeneracy and quasiparticle excitations that have restricted motion.
These phases have been the subject of much research in recent years (see the recent review, Ref.~\onlinecite{Nandkishore2018-ee}, and references within). 
It has been discovered that fracton theories arise naturally as a result of a generalized gauging procedure~\cite{Vijay2016-dr,Williamson2016-lv,Shirley2018-en,You2018-as} applied to systems with subsystem symmetries.
For example, the plaquette Ising model, when gauged, results in the X-cube model of fracton order~\cite{Vijay2016-dr}.

It has also been recently appreciated that subsystem symmetries, like global symmetries, are also capable of protecting non-trivial SPT phases,
 called subsystem SPT (SSPT) phases~\cite{You2018-em}.
The first discovered example arose in the context of quantum information, where it was found that the square lattice cluster model~\cite{Raussendorf2001-xm}, if protected by a set of line-like subsystem symmetries, could act as a universal resource for measurement-based quantum computation~\cite{Else2012-li,Raussendorf2018-nh} (MBQC) throughout the phase, a result which has recently also been extended to other SSPT phases~\cite{Devakul2018-di,Stephen2018-qn} (we note that an interesting approach to defining subsystem symmetries from an underlying quantum cellular automaton is introduced in Ref.~\onlinecite{Stephen2018-qn}).
In 1D, the computational power of an SPT phase under MBQC has been shown to be directly related to their classification~\cite{Else2012-ie,Else2012-li,Miller2015-gl,Stephen2017-cp,Raussendorf2017-gb}.
However, MBQC is only universal in dimension two or higher, and an extension of such results to SSPTs is not straightforward.
For one, there is currently no theory of classification for SSPTs.
In Ref.~\onlinecite{You2018-em}, the term ``weak'' SSPT was used to describe SSPTs that were essentially composed of decoupled 1D SPT chains, as opposed to a ``strong'' SSPT, such as the square lattice cluster model which is not constructed of decoupled 1D SPT chains.
It was noted~\cite{Devakul2018-di,Stephen2018-qn} that this example of a weak SSPT could not serve as a resource for universal MBQC using only single-spin measurements.
However, it is not clear how general this statement is as there is currently no clear definition for what specifies a strong or weak SSPT.
In this paper, we hope to tackle the question of what constitutes a weak or strong SSPT, and whether such SSPTs may be classified in a natural way.

Our work draws inspiration from a series of recent works on fracton topological orders, where the concept of a \emph{foliated fracton phase} has been introduced~\cite{Shirley2017-fi,Shirley2018-bx,Shirley2018-jy,Shirley2018-yj,Shirley2018-en} to classify non-fractal (Type-I~\cite{Vijay2016-dr}) fracton orders.
A foliated fracton phase is an equivalence class of fracton topological orders, whereby two fracton phases are considered equivalent if one can be brought to the other via a combination of local unitary~\cite{Chen2010-sm} (LU) evolution and the addition and removal of 2D \emph{topologically ordered phases}.
We remark that standard phase equivalence only allows the addition of trivial product states along with LU evolution. Therefore, foliated fracton phases present a drastic departure from the norm.
Foliated fracton phases may be thought of as a 3D phase equivalence ``modulo'' any 2D physics: this motivates a similar construction for SSPTs.

We propose a natural definition of a strong equivalence relation for two-dimensional SSPTs protected by line-like symmetries,
whereby two phases are in the same equivalence class if they can be connected to each other via a \emph{linearly-symmetric local unitary} (LSLU) evolution, which we will define.
By construction, the weak SSPT composed of decoupled 1D SPT chains may be transformed into a trivial product state via an LSLU evolution.
Importantly, we find that the square lattice cluster model \emph{cannot} be transformed to the trivial state.
We may therefore take this equivalence relation to \emph{define} a strong SSPT phase as one that cannot be connected to the trivial product state via an LSLU evolution.
Moreover, we find that there are several distinct equivalence classes of strong SSPTs, which are in one-to-one correspondence with the non-trivial elements of the group
\begin{equation}
\label{eq:cgs}
    \mathcal{C}[G_s] = \mathcal{H}^2[G_s^2,U(1)]/(\mathcal{H}^2[G_s,U(1)])^{3}
\end{equation}
where $G_s$ is the finite abelian onsite symmetry group characterizing the subsystem symmetries (to be defined), and $\mathcal{H}^2[G,U(1)]$ is the second cohomology group which classifies the projective representations of $G$. We have utilized the fact that $(\mathcal{H}^2[G_s,U(1)])^{3}$ always appears as a (normal) subgroup of $\mathcal{H}^2[G_s^2,U(1)]$ for abelian $G_s$ (for details see section~\ref{subsec:classification}).
This therefore presents a classification for strong SSPT phases, according to our strong phase equivalence.
Finally, we note that the equivalence class defined by LSLU is the same as that defined by standard symmetric local unitaries in combination with stacking with 1D SPT chains (See Sec~\ref{sec:stacking}), which indeed has a natural interpretation of being a 2D equivalence class ``modulo'' 1D physics.

In an appendix of Ref.~\onlinecite{You2018-em}, it was argued that strong SSPT phases did not exist for conventional continuous symmetry groups such as $U(1)$ or $SU(2)$, while the effect of an additional global $\mathbb{Z}_2^{T}$ time reversal symmetry does not lead to new strong phases, as diagnosed by the projective representation at the edge.
Furthermore, a non-abelian $G_s$ implies the existence of a local symmetry, as we will show.
We therefore focus on unitary representations of finite abelian groups $G_s$, which encompass most known examples of strong SSPTs (e.g. $\mathbb{Z}_2$ or $\mathbb{Z}_n\times\mathbb{Z}_m$).

In Sec.~\ref{sec:setting}, we define the class of models we are interested in, and what we mean by subsystem symmetries.
Sec.~\ref{sec:standard} contains a review of standard 1D SPT phase equivalence and classification, in addition to a review of various useful tools such as projective representations.  
Then, in Sec.~\ref{sec:sspt} we present our strong phase equivalence and classification of strong SSPTs, for our general class of models.
Sec.~\ref{sec:examples} then walks through the results of the previous section with an example at hand, the square lattice cluster model.
Finally, we finish with a few additional comments and conclusions in Sec.~\ref{sec:aspects} and~\ref{sec:conclusion}.
These include some straightforward generalizations as well as a connection to spurious topological entanglement entropy~\cite{KitaevPreskill,levinwenentanglement} observed in non-topologically ordered states on a cylinder~\cite{PhysRevB.94.075151}.

\section{Setting}\label{sec:setting}
Let us consider 2D models protected by line-like subsystem symmetries of a specific form.
Let bosonic degrees of freedom live at the sites $s$ of a square lattice, with local Hilbert space $\mathcal{H}_{xy}$ at site $(x,y)$,
such that the total Hilbert space is $\mathcal{H}=\bigotimes_{xy} \mathcal{H}_{xy}$.
Each site transforms as a unitary linear representation of some \emph{onsite} symmetry group, $G_s$, which we take to be finite and abelian.
For each element $g \in G_s$, we demand that the system respects the following \emph{subsystem} symmetries,
\begin{align}
\begin{split}
S^{v}_{x}(g) &= \prod_{y=-\infty}^{\infty} u_{x y}(g)\\
S^{h}_{y}(g) &= \prod_{x=-\infty}^{\infty} u_{x y}(g)
\end{split}\label{eq:symdef}
\end{align}
for every $x,y\in\mathbb{Z}$,
where $u_{xy}(g)$ is the on-site unitary (faithful) representation transforming the site $(x,y)$ by $g$. 
$S^{v}$ and $S^h$ act along vertical and horizontal rows, respectively, as illustrated in Figure~\ref{fig:setting}.
The \emph{total} symmetry group is therefore a (sub)extensively large group (which should not be confused with the finite onsite symmetry group $G_s$).
We consider local short-range entangled Hamiltonians which respect all these symmetries. 

We remark that if $G_s$ were a non-abelian group, then the symmetry 
\begin{equation}
S_y^{h}(g_2^{-1}) S_x^{v}(g_1^{-1}) S_y^{h}(g_2)  S_x^v(g_1)  = u_{xy}(g_2^{-1} g_1^{-1} g_2 g_1)
\end{equation}
may act non-trivially on only a single site if ${g_2^{-1} g_1^{-1} g_2 g_1\neq 1}$.
This implies the existence of a local $[G_s,G_s]$ symmetry on every site, and an effective abelian ${G_s^\prime=G_s/[G_s,G_s]}$ subsystem symmetry.
We therefore focus our attention on abelian groups $G_s$ from the beginning. 
We also note that Eq.~\eqref{eq:symdef} induces an identification of the group elements $g\in G_s$ across all sites of the system. 

These symmetries present a drastic change from the now well-understood phases protected by a global on-site symmetry group $G_s$ in 2D, which are classified by the 3rd cohomology group $\mathcal{H}^3[G_s,U(1)]$.
What distinct phases are possible under such subsystem symmetries?  

Consider the following scenario: suppose we construct a 2D phase by aligning 1D SPT chains horizontally, in such a way that all the vertical symmetries are still respected (in this process a single SPT chain may span multiple rows in order to respect all the vertical symmetries).
We call such a phase a ``weak'' SSPT~\cite{You2018-em}.  
Under the standard SPT phase equivalence, which we will review briefly in Sec.~\ref{sec:standard}, two states are in the same phase if they can be adiabatically transformed to one another while respecting the symmetry, via a symmetric local unitary (SLU) evolution.
In our weak SSPT, each 1D SPT chain could be in any allowed 1D SPT phase, and by this definition these are all distinct phases.
The number of distinct phases therefore grows exponentially with the system size.
Note that we never assume any translational invariance in any of our discussion.
Nevertheless, we would like to be able to make a clear distinction between these weak SSPT phases and a ``strong'' SSPT phase, which cannot be written as a product of 1D SPT phases.

To this end, the main result of this paper is a definition of a strong equivalence relation for SSPT phases in Sec.~\ref{sec:sspt}, under which all weak SSPT phases are equivalent to the trivial phase.
This \emph{defines} the meaning of a strong SSPT phase.
Our secondary result is a classification of strong SSPT phases under this equivalence relation: strong SSPT phases may be classified according to the group $\mathcal{C}[G_s]$ in Eq.\eqref{eq:cgs}.
As an example, in Sec.~\ref{sec:examples} we show that the SSPT phase of the square lattice cluster model, which has the onsite symmetry group $G_s=\mathbb{Z}_2\times\mathbb{Z}_2$, is non-trivial under this equivalence relation.
We further show that these equivalence classes of strong SSPT phases are in one-to-one correspondence with elements of the group $\mathcal{C}[(\mathbb{Z}_2)^2]=(\mathbb{Z}_2)^6$, and exhibit the group structure under stacking.

\section{Standard SPT phase equivalence}\label{sec:standard}
To set the stage for our discussion of 2D SSPT phases, we first present a review of the relevant standard concepts coming from the study of 1D SPT phases.

\subsection{Symmetric local unitary transformations}\label{sec:slu}
Let $\ket{\psi}$ be the unique ground state of a gapped local Hamiltonian $H$, with symmetry group $G$ with an onsite representation $u_x(g)$ on site $x$.
The symmetries are $S(g) = \prod_x u_x(g)$ for $g\in G$, and the Hamiltonian respects $[S(g),H]=0$.
Two states $\ket{\psi}$ and $\ket{\psi}^\prime$ are said to be in the same SPT phase if there exists a symmetric local unitary (SLU) evolution, $U_\mathrm{SLU}$, that connects the two: $\ket{\psi} = U_\mathrm{SLU} \ket{\psi}^\prime$.
A state is in the trivial phase if it can be connected via an SLU to a product state.
For convenience, we may always express an SLU evolution as a \emph{symmetric finite-depth quantum circuit}, which we now define.

A quantum circuit of depth $d$ representing an SLU evolution, $U_\mathrm{SLU}$, may be represented as
\begin{equation}
    U_\mathrm{SLU} = U_{pw}^{(d)}U_{pw}^{(d-1)}\dots U_{pw}^{(1)}.
\end{equation}
Here, each $U_{pw}$ is a \emph{piecewise local} unitary operator, 
\begin{equation}
    U_{pw} = \bigotimes_{i} U^{(i)}_{s} 
\end{equation}
where $\{U_{s}^{(i)}\}$ are \emph{local symmetric} unitary operators which all act on local \emph{disjoint} regions.
Importantly, the radius of support for each $U_{s}$ must be bounded by some finite length. 
Finally, to represent an SLU evolution, we require that $[S(g),U_{s}]=0$ for all $U_{s}$.
Without this symmetry restriction, all short-range entangled phases can be connected to a product state (via an LU).
Such a quantum circuit is shown in Fig.~\ref{fig:slu} for a 1D chain.
Two quantum states are in the same SPT phase if and only if there exists a quantum circuit $U_\mathrm{SLU}$ connecting the two, where $d$ is a finite constant.

\subsection{Projective Representations}
In 1D, SPT phases with symmetry group $G$ are in one-to-one correspondence with the \emph{projective representations} of the group $G$~\cite{Chen2013-gq}.  
Projective representations will also play a key role in our classification of strong SSPT phases, so we present an introduction here.
A non-projective or \emph{linear} representation of a group $G$ is a mapping from group elements $g\in G$ to unitary matrices $V(g)$, such that $V(g_1)V(g_2) = V(g_1 g_2)$, for all $g_1,g_2\in G$.
A representation $V$ is \emph{projective} if it instead satisfies
\begin{equation}
    V(g_1) V(g_2) = \omega(g_1,g_2) V(g_1 g_2),
\end{equation}
where $\omega(g_1,g_2)\in U(1)$ is a phase, referred to as the factor system of a particular projective representation.
The factor system must satisfy
\begin{align}
    \omega(g_1,g_2)\omega(g_1g_2,g_3) &= \omega(g_1,g_2g_3)\omega(g_2,g_3)\label{eq:factorsystem}\\
    \omega(1,g_1) = \omega(g_1,1) &= 1
\end{align}
for all $g_1,g_2,g_3\in G$, where $1$ is the identity element.

A different choice of prefactors, $V^\prime (g) = \alpha(g) V(g)$, leads to the factor system 
\begin{equation}
    \omega^\prime(g_1,g_2) = \frac{\alpha(g_1g_2)}{\alpha(g_1)\alpha(g_2)}\omega(g_1,g_2)
    \label{eq:factorsystemequiv}
\end{equation}
Two factor systems $\omega(g_1,g_2)$ and $\omega^\prime(g_1,g_2)$ related in this way are said to be equivalent, and both belong to the same equivalence class $\omega$.  

Given two projective representations $V_1(g)$ and $V_2(g)$ with factor systems $\omega_1(g_1,g_2)$ of equivalence class $\omega_1$, and $\omega_2(g_1,g_2)$ of equivalence class $\omega_2$, we may define the projective representation
\begin{equation}
    V(g) = V_1(g)\otimes V_2(g)
\end{equation}
with factor system
\begin{equation}
    \omega(g_1,g_2) = \omega_1(g_1,g_2)\omega_2(g_1,g_2)
\end{equation}
which now belongs to the class $\omega$, defining a group operation $\omega_1\omega_2 = \omega$.
Under this operation, the equivalences classes form an abelian group which is given by the second cohomology group $\mathcal{H}^2[G,U(1)]$.
The identity element of $\mathcal{H}^2[G,U(1)]$ corresponds to the linear representations, while other elements correspond to non-trivial projective representations.

We consider cases where $G$ is a finite abelian group.  
In this case, projective representations simply allow for non-trivial commutation relations of the form
\begin{equation}
    V(g_1) V(g_2) =  \phi(g_1,g_2) V(g_2) V(g_1)
\end{equation}
where 
\begin{equation}
    \phi(g_1,g_2) = \omega(g_1,g_2)/\omega(g_2,g_1)
\end{equation}
is explicitly invariant under equivalence transformations of the form in Eq.~\ref{eq:factorsystemequiv}, and can therefore be regarded as a signature of the class $\omega$.  
Under the group operation on two classes, $\omega = \omega_1\omega_2$, we have that $\phi(g_1,g_2) = \phi_1(g_1,g_2)\phi_2(g_1,g_2)$.

As an example, consider the group 
\begin{equation}
    G=\mathbb{Z}_2\times \mathbb{Z}_2=\{1,g_a,g_b,g_a g_b\}
\end{equation}
where $g_a$ and $g_b$ are defined to be the two generators for $G$.  
In this case, there are two classes of projective representations: the trivial linear representation where $\phi(g_a,g_b)=1$, and the non-trivial projective representation with $\phi(g_a,g_b)=-1$.
An example of the latter is given by the Pauli representation, 
\begin{align}
    V(g_a) = X,\ V(g_b) = Z,\ V(g_a g_b) = XZ
\end{align}
where $X$,$Z$, are the Pauli matrices,
with non-trivial $\omega(g_1,g_2)$ given by
\begin{align}
    \omega(g_b,g_a) = \omega(g_a g_b,g_a) = \omega(g_a g_b, g_a g_b) =  -1
\end{align}
In this case $\phi$ is a complete invariant and the projective representations of $\mathbb{Z}_2\times\mathbb{Z}_2$ are therefore in one-to-one correspondence with elements in $\mathcal{H}^{2}[\mathbb{Z}_2\times\mathbb{Z}_2,U(1)] = \mathbb{Z}_2$.

\subsection{1D classification}
Non-trivial SPT phases in 1D may be identified by their non-trivial edges, where the symmetry group $G$ is realized projectively leading to a symmetry-protected degeneracy at the edge.
We motivate this classification in a way that will prove useful for our classification of strong SSPT phases to follow.

Let  $\ket{\psi}$ be the ground state of a gapped symmetric local Hamiltonian in the absence of a boundary, with symmetry group $G$ which we take to be finite and abelian (as these are the ones relevant for the case of SSPTs). 
As the ground state is unique, we must have ${\ket{\psi} = S(g)\ket{\psi}}$ up to a phase which can be absorbed into $S(g)$.
Now, consider the truncated symmetry operator 
\begin{equation}
    U_{[x_0,x_1)}(g) = \prod_{i=x_0}^{x_1-1} u_i(g)
\end{equation}
where $x_0<x_1$ are the endpoints, which we take to be separated by much further than the correlation length.
Acting on $\ket{\psi}$, this may create two local excitations in the neighborhood of $x_0$ and $x_1$.  
These excitations may be locally annihilated by some unitary operators $V^{L}_{x_0}(g)$ and $V^{R}_{x_1}(g)$ with support size on the order of the correlation length about $x_0$ and $x_1$, such that
\begin{equation}
    V^{L}_{x_0}(g) V^{R}_{x_1}(g) U_{[x_0,x_1)}(g) \ket{\psi} = \ket{\psi}.\label{eq:1dvvs}
\end{equation}
Note that in writing this, we have assumed that $\ket{\psi}$ is short-range entangled and not spontaneous symmetry breaking, as is the case for SPT phases.

We may also simplify this picture by not distinguishing between local excitations created at the left/right endpoints of $U(g)$, as they can be related to each other by a symmetry operation:
\begin{equation}
     U_{[x_0,x_1)}(g) S(g^{-1}) = U_{[-\infty,x_0)}(g^{-1}) U_{[x_1,\infty)}(g^{-1})
\end{equation}
Thus, the local excitation created at the right (left) end of $U(g)$ is the same as the excitation created at the left (right) end of $U(g^{-1})$.
We may therefore simply substitute $V_{x}(g) \equiv V_{x}^{L}(g)$ on the left endpoint and $V_{x}(g^{-1})$ on the right endpoint,
and choose a $U(1)$ phase prefactor such that Eq.~\ref{eq:1dvvs} is still satisfied.



Observe the commutation relation,
\begin{equation}
     \left[S(g_1),  V_{x_0}(g_2) V_{x_1}(g_2^{-1}) U_{[x_0,x_1)}(g_2)\right] \ket{\psi}  = 0
     \label{eq:1dcommrel}
\end{equation}
While $[S(g_1), U_{[x_0,x_1)}(g_2)]=0$ as the onsite representation $u_i(g)$ is linear, $S(g_1)$ need not commute with $V_{x_0}(g_2)$ and $V_{x_1}(g_2^{-1})$ individually.  
Indeed, we may have that when acting on $\ket{\psi}$,
\begin{equation}
\begin{split}
    S(g_1) V_{x_0}(g_2)\ket{\psi} &= \phi^{*}(g_1,g_2) V_{x_0}(g_2) S(g_1) \ket{\psi}\\
    S(g_1) V_{x_1} (g_2^{-1})\ket{\psi} &= \phi(g_1,g_2) V_{x_1} (g_2^{-1}) S(g_1) \ket{\psi}
\end{split}\label{eq:1dphis}
\end{equation}
which still satisfies Eq.~\ref{eq:1dcommrel}, where $\phi(g_1,g_2)$ is a $U(1)$ phase.
Note that this phase cannot depend on $x_0$, $x_1$, nor on choice of $V$s, as we may change each independently of the others --- it is therefore a global property of the bulk.

Now suppose we introduce edges into the system at $x=1$ and $x=\ell$.  
As the ground state need not be unique in the presence of an edge, we no longer require that $S(g)\ket{\psi} = \ket{\psi}$ (it may move $\ket{\psi}$ around in the ground state manifold).  
Instead, we may find local operators $V_1(g)$ and $V_\ell(g^{-1})$ on the edges such that
\begin{equation}
    V_{1}(g) V_{\ell}(g^{-1}) S(g) \ket{\psi} = \ket{\psi}
\end{equation}
for any $\ket{\psi}$ in the ground state manifold.
Put differently, this means that
\begin{equation}
    S(g) \ket{\psi} = V^{\dagger}_{1}(g) V^{\dagger}_{\ell}(g^{-1}) \ket{\psi}
\end{equation}
and we may decompose $S(g)$ into separated operations with support at the left and right edges separately,
which act nontrivially only within the ground state manifold.
Repeating our previous analysis with $x_0$ or $x_1$ at an edge, we find that the representation of the symmetry on the left edge, $V_\mathrm{e}(g)\equiv V_1^{\dagger}(g)$, may be projective, and the class is completely determined by the previously discovered bulk $\phi$, 
\begin{equation}
    V_\mathrm{e}(g_1) V_\mathrm{e}(g_2) = \phi(g_1,g_2) V_\mathrm{e}(g_2) V_\mathrm{e}(g_1)
\end{equation}
Similarly, operators at the right edge must exhibit the same projective representation.
A non-trivial projective representation requires a degenerate ground space manifold on which the matrices $V_\mathrm{e}$ may act, thus leading to the protected edge modes of non-trivial SPTs.

Phases with projective representation from different classes cannot be adiabatically transformed into one another via a SLU evolution. While such an evolution may change $V$ in Eq.~\eqref{eq:1dphis}, it must leave $\phi$ invariant. 
Meanwhile, if two phases are of the same class, then there is no obstacle to connecting the two adiabatically.
By the phase equivalence relation given in Sec.~\ref{sec:slu}, distinct SPT phases are in one-to-one correspondence with the projective representations of $G$, and can be diagnosed by the projective representation observed at the edges~\cite{Chen2011-ss,Else2014-ar}.


Now consider two states, $\ket{\psi_1}$ and $\ket{\psi_2}$, characterized by the projective classes $\omega_1$ and $\omega_2$.
Consider the 1D system obtained by stacking the two chains on top of each other, such that $\ket{\psi}=\ket{\psi_1}\otimes\ket{\psi_2}$, with the symmetry acting onsite as $u_i(g)=u_{i,1}(g)\otimes u_{i,2}(g)$.
Following the above, the projective action of the symmetry at the left edge is given by $V_e(g) = V_{e,1}(g)\otimes V_{e,2}(g)$, which is of the class $\omega=\omega_1\omega_2$.
Therefore, under stacking, SPT phases form a group structure given by the second cohomology group $\mathcal{H}^{2}[G,U(1)]$.


\begin{figure}
    \centering
\includegraphics[width=0.4\textwidth]{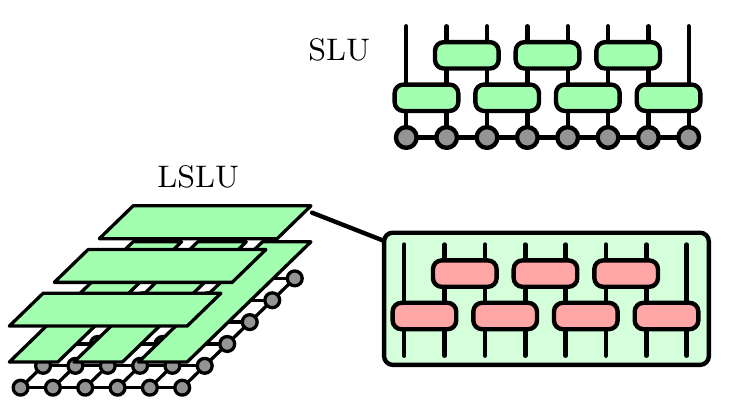}
    \caption{A symmetric local unitary (SLU) evolution expressed as a quantum circuit defines phase equivalence in SPTs protected by global symmetries.
    Each green gate represents a symmetry-respecting unitary evolution.
    We propose to define a strong equivalence relation for 2D SSPT phases using linearly-symmetry local unitary evolution, in which each individual local gate (shown in red) need not be symmetric, but must be grouped into gates acting along straight lines which are, as a whole, symmetric.
    }
    \label{fig:slu}
\end{figure}

\section{Strong equivalence of SSPT phases}\label{sec:sspt}
Having set the stage with a review of 1D SPT phase equivalence, we now turn to our main topic of interest: strong SSPTs in 2D.
Recall that we take such a model to be defined with respect to an onsite finite abelian symmetry group $G_s$, which in turn defines the total symmetry group generated by the set of $S^v_{x}(g)$ operators, which act along vertical columns with fixed $x$, and the set of $S^h_{y}(g)$ operators, which act along horizontal rows with fixed $y$, for $g\in G_s$.

\subsection{Linearly symmetric local unitary transformations}

To proceed, we introduce the concept of a \emph{linearly symmetric local unitary} (LSLU) evolution, which are a generalization of the previously defined SLU evolution, and take the form shown in Fig.~\ref{fig:slu} (bottom).
Such an evolution may be constructed as a finite-depth quantum circuit $U_{LSLU}$,
\begin{equation}
    U_\mathrm{LSLU} = U_{lpw}^{(d)}U_{lpw}^{(d-1)}\dots U_{lpw}^{(1)}
\end{equation}
where each $U_{lpw}$ is a \emph{linearly piecewise} unitary, taking the form
\begin{equation}
    U_{lpw} = \bigotimes_{i} U_{ls}^{(i)}
\end{equation}
where $\{U_{ls}^{(i)}\}$ are \emph{linearly-supported symmetric} local unitaries with disjoint support.
By linearly-supported, we mean that the support of $U_{ls}$ may extend indefinitely in either the $x$ or $y$ direction, but only a small finite range in the other.  
A single green rectangle in Fig.~\ref{fig:slu}~(bottom) represents one $U_{ls}$.
We also require that they all commute with all symmetries,
\begin{equation}
    [U_{ls}, S^v_x(g)] = [U_{ls}, S^h_y(g)] = 0
\end{equation}
for all $x$, $y$, and $g\in G_s$.

The only restriction on $U_{ls}$ beyond this is that it must be a local unitary transformation.  
For completeness, we may express $U_{ls}$ as a finite depth $\delta$ quantum circuit.
\begin{equation}
    U_{ls} = U_{pw}^{(\delta)} U_{pw}^{(\delta-1)} \dots U_{pw}^{(1)}
\end{equation}
where each $U_{pw}$ is a piecewise local unitary operator, given by
\begin{equation}
    U_{pw} = \bigotimes_i U_\mathrm{loc}^{(i)}
\end{equation}
where $\{U_\mathrm{loc}^{(i)}\}$ are disjoint unitary operators with a finite radius of support.
Crucially, neither any $U_\mathrm{loc}$ nor $U_{pw}$ need respect any symmetries ---   
only the final product, $U_{ls}$, need respect all subsystem symmetries.
The total depth of the circuit is given by $d \delta$, and must be a constant independent of system size for it to represent an LSLU.

Conceptually, an SLU may be represented as a quantum circuit where each gate must respect all symmetries.
In an LSLU, each gate need not individually respect the symmetries, but there must be a way of grouping the gates into disjoint operations acting along vertical or horizontal lines such that the combined action along the line as a whole respects the symmetries.

The first main result of this paper is the proposal of the following equivalence relation:
\emph{
Two SSPTs, with unique ground states $\ket{\psi}$ and $\ket{\psi^\prime}$, are in the same \emph{strong SSPT equivalence class} if there exists a finite-depth LSLU circuit $U_\mathrm{LSLU}$ connecting the two, such that $\ket{\psi}=U_\mathrm{LSLU} \ket{\psi^\prime}$.
}

The motivation for this definition comes from the observation that any 1D SPT may be deformed to a product state via a (non-symmetric) LU evolution.
If both the initial and final states are symmetric, we may take this LU to be, as a whole, symmetric.  
Consider a weak SSPT phase consisting of 1D chains aligned horizontally.  
An LU along the horizontal direction is able to disentangle a 1D chain, while remaining symmetric as a whole.  
Such an operation is allowed in an LSLU, and are represented by $U_{ls}$ above.
Thus, by allowing an LSLU in our equivalence relation, we are essentially ``modding out'' 1D chains.  
Whatever remains must contain some fundamentally two-dimensional physics.
This is similar in spirit to the definition of foliated~\cite{Shirley2018-bx,Shirley2018-jy,Shirley2018-yj,Shirley2018-en,Shirley2017-fi} fracton phases, where the equivalence relation for 3D foliated fracton phases is defined modulo the addition or removal of 2D topological orders.

An example of an LSLU is shown in Figure~\ref{fig:lsluexample}, for the explicit case of the square lattice cluster model.
This LSLU consists of a product of controlled-Z gates which as a whole commutes with all subsystem symmetries.

The goal now is to show that there indeed exists non-trivial equivalence classes under this definition, which leads to a classification of such phases.

\subsection{Bulk Invariants}\label{ssec:bulk}
In this section, we derive the existence of bulk properties (much like the projective phases $\phi(g_1,g_2)$ in the 1D SPT classification) that are invariant under LSLU transformations.
Later, in Section~\ref{sec:examples}, we give an explicit example which makes this construction clear.

First, let us introduce the truncated symmetry operation, 
\begin{equation}
    U_{x_0x_1}^{y_0y_1}(g) = \prod_{x=x_0}^{x_1-1} \prod_{y=y_0}^{y_1-1} u_{xy}(g)
\end{equation}
where $x_0<x_1$, $y_0<y_1$, for any $g\in G_s$, which represents the application of the symmetry $g$ to a rectangular region of the system.
Let us take $|x_1-x_0|$ and $|y_1-y_0|$ to be much larger than any correlation length in the system.

\begin{figure}
    \centering
\includegraphics[width=0.5\textwidth]{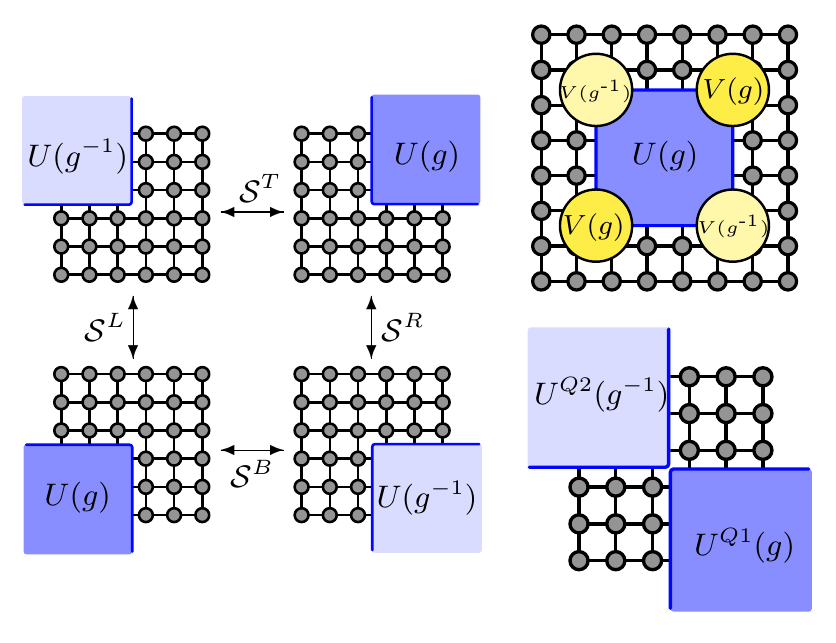}
    \caption{
    A rectangular truncated symmetry operation $U_{x_0 x_1}^{y_0 y_1}(g)$ creates local excitations only at its \emph{corners}.
    (left) Corners of different orientations that can be related to one another other by the action of symmetries, as shown here, by applying $\mathcal{S}(g)$ or $\mathcal{S}(g^{-1})$ of different orientations. 
    (top right) The four excitations created by $U_{x_0 x_1}^{y_0 y_1}(g)$ may be locally annihilated by operators $V_{xy}(g)$ or $V_{xy}(g^{-1})$ at its corners.
    (lower right) We show the symmetry operation $\mathcal{S}^Q_{xy}(g)$ discussed in the text.
    }
    \label{fig:corners}
\end{figure}

We may think of $U^{y_0y_1}_{x_0x_1}(g)$ as the application of $S^v_{x}(g)$ truncated to $[y_0,y_1)$, for $x\in[x_0,x_1)$.
We may also alternatively think of it as the application of $S^h_{y}(g)$ truncated to $[x_0,x_1)$, for $y\in[y_0,y_1)$.
Therefore, the only place where $U^{y_0y_1}_{x_0x_1}(g)$ does not look like the application of a symmetry is near its corners, and it may therefore create four local excitations at each of the corners.
As the ground state is unique, short-range entangled, and symmetric, these local excitations may be locally annihilated via a unitary operator at each of the corners, such that
\begin{equation}
    V_{x_0y_0}^{BL}(g)
    V_{x_0y_1}^{TL}(g)
    V_{x_1y_0}^{TR}(g)
    V_{x_1y_1}^{BR}(g)
    U^{y_0y_1}_{x_0x_1}(g) \ket{\psi} = \ket{\psi}\label{eq:rect1}
\end{equation}
Here, $B(T)L(R)$ indicates the bottom (top) left (right) corner of the rectangle, and $V_{xy}(g)$ only has support within a correlation-length sized region near $(x,y)$.
Each $V_{xy}(g)$ must form some (possibly projective) representation of $G_s$, which may depend on our choice of $V$s.
In Figure~\ref{fig:corners}~(left), we show how different types of corners may be related to each other by application of subsystem symmetries: we therefore do not need to distinguish between them and we may simply define 
\begin{equation}
V_{xy}(g) \equiv V_{xy}^{BL}(g) = V_{xy}^{TL}(g^{-1}) =  V_{xy}^{TR}(g) = V_{xy}^{BR}(g^{-1})
\end{equation}
with a choice of overall phase prefactors such that Eq.~\ref{eq:rect1} becomes
\begin{equation}
    V_{x_0y_0}(g)
    V_{x_0y_1}(g^{-1})
    V_{x_1y_1}(g)
    V_{x_1y_0}(g^{-1})
    U^{y_0y_1}_{x_0x_1}(g) \ket{\psi} = \ket{\psi}\label{eq:rect2}
\end{equation}
as illustrated in Figure~\ref{fig:corners}~(right).
The operators $V_{xy}(g)$ contain all the information we need about the system.

Let us define the symmetry operations,
\begin{align}
    \mathcal{S}^{R}_{x}(g) &= \prod_{x^\prime = x}^{\infty} S^{v}_{x^\prime}(g),\hspace{0.5cm}
    \mathcal{S}^{L}_{x}(g) = \prod_{x^\prime = -\infty}^{x-1} S^{v}_{x^\prime}(g)\\
    \mathcal{S}^{T}_{y}(g) &= \prod_{y^\prime = y}^{\infty} S^{h}_{y^\prime}(g),\hspace{0.5cm}
    \mathcal{S}^{B}_{y}(g) = \prod_{y^\prime = -\infty}^{y-1} S^{h}_{y^\prime}(g)
\end{align}
where the superscript denotes that we are applying the symmetry $g$ to all sites to the right, left, top, or bottom of the coordinate $x$ or $y$.
We will mainly use $\mathcal{S}^{R}$ and $\mathcal{S}^{T}$, although we have defined them all for completeness.

We now proceed to prove that 
the $U(1)$ phase $\beta_{xy}^{R}(g)$, given by
\begin{align}
\begin{split}
    \beta_{xy}^{R}(g) &= \bra{\psi} \mathcal{S}^{\dagger R}_{x}(g)V_{xy}^{\dagger}(g) \mathcal{S}^{R}_{x}(g) V_{xy}(g)\ket{\psi} 
    \label{eq:betadef}
\end{split}
\end{align}
\emph{is independent of $x$, $y$, and therefore cannot be changed by an LSLU evolution.}
We could have also chosen any orientation, $R$, $T$, $L$, or $B$, which would give the same value.
We may therefore simply refer to $\beta(g) = \beta_{xy}^{R}(g)$.
Note that $\beta(g)$ is nothing but the phase obtained from commuting $S_{xy}^{R}(g)$ with $V_{xy}(g)$, when acting on the ground state. 

The proof consists of three steps: 
First, we prove that $\beta^{R}_{xy}(g)$ is the same for all $y$, and then that $\beta^{T}_{xy}(g)$ is the same for all $x$.
Then, we show that for a given $(x,y)$, $\beta^{R}_{xy}(g) = \beta^{T}_{xy}(g)$.
It then follows that $\beta(g) = \beta_{xy}^{R}(g)$ is independent of $x$,$y$.

The first two steps can be accomplished by looking at our rectangle operator $U_{x_0x_1}^{y_0y_1}(g)$ from earlier.
Since the combination in Eq.~\ref{eq:rect2} acts trivially on $\ket{\psi}$, as does $\mathcal{S}^{R/T}_{x_1}(g)$, they must commute when acting on $\ket{\psi}$. 
Let us first deal with $R$.
$\mathcal{S}^{R}_{x_1}(g)$ only overlaps with $V$s from the top right and bottom right corners: $V_{x_1 y_0}(g^{-1})$ and $V_{x_1 y_1}(g)$.  
Therefore, 
\begin{equation}
    \left[\mathcal{S}^{R}_{x_1}(g), V_{x_1 y_0}(g^{-1}) V_{x_1 y_1}(g) \right]\ket{\psi} = 0
\end{equation}
but we also have that, from our definition of $\beta_{x y}^{R}(g)$,
\begin{align}
\begin{split}
    \mathcal{S}^{R}_{x_1}(g) V_{x_1 y_1}(g)\ket{\psi} &= \beta_{x_1 y_1}^{R}(g) V_{x_1 y_1}(g) \mathcal{S}^{R}_{x_1}(g)\ket{\psi} \\
    \mathcal{S}^{R}_{x_1}(g) V_{x_1 y_0}(g^{-1})\ket{\psi} &= \beta_{x_1 y_0}^{*R}(g) V_{x_1 y_0}(g^{-1}) \mathcal{S}^{R}_{x_1}(g)\ket{\psi} 
\end{split}
\end{align}
(note that $V_{xy}(g^{-1}) = \omega_{xy}(g,g^{-1})V^\dagger_{xy}(g)$, where $\omega_{xy}(g,g^\prime)$ is the factor system of the representation $V_{xy}(g)$).
Therefore, we must have that
\begin{equation}
\beta_{x_1 y_1}^{R} = \beta_{x_1 y_0}^{R} 
\end{equation}
However, we could have chosen $y_0$ or $y_1$ independently.  
This phase $\beta_{x y}^{R}(g)$ is therefore independent of $y$.
Similarly, from $T$, we discover that $\beta_{x y}^{T}(g)$ must be independent of $x$.

The final step requires another ingredient.
Consider the symmetry operation
\begin{equation} 
    \mathcal{S}^{Q}_{xy}(g) = \mathcal{S}_{x}^{R}(g) [\mathcal{S}_{y}^{T}(g)]^\dagger
    = \mathcal{S}_{x}^{R}(g) \mathcal{S}_{y}^{T}(g^{-1})
\end{equation}
which consists of applying $g$ in the bottom right quadrant, and $g^{-1}$ in the top left quadrant (it acts as identity on the top right quadrant), as shown in Figure~\ref{fig:corners} (lower right).

We now show that $\mathcal{S}^{Q}_{xy}(g)$ commutes with $V_{xy}(g)$.  
To do this, notice that we may split $\mathcal{S}^{Q}_{xy}$ as 
\begin{align} 
    \mathcal{S}^{Q}_{xy}(g) &= 
    \left[\prod_{x^\prime=x}^{\infty}\prod_{y^\prime=-\infty}^{y-1} u(g)\right] 
    \left[ \prod_{x^\prime=-\infty}^{x-1}\prod_{y^\prime=y}^{\infty} u(g^{-1})\right]
    \\
    &\equiv \left[ U^{Q2}_{xy}(g) \right]\left[U^{Q1}_{xy}(g^{-1})\right]
\end{align}
where $U^{Q1}_{xy}(g^{-1})$ only has support in the top left quadrant, and $U^{Q2}_{xy}(g)$ in the bottom right quadrant, about $(x,y)$.
Importantly, they only touch each other at the point $(x,y)$, as shown in Figure~\ref{fig:corners} (lower right).
Then, supposing that 
\begin{equation}
V_{xy}(g^{-1}) U_{xy}^{Q1}(g)\ket{\psi}= V_{xy}(g^{-1}) U_{xy}^{Q2}(g)\ket{\psi} = \ket{\psi} \label{eq:dangerous}
\end{equation}
and using $U_{xy}^{Q1}(g) = [U_{xy}^{ Q1}(g^{-1})]^{\dagger}$,
we have that
\begin{align}
\begin{split}
    & V_{xy}^{\dagger}(g) \mathcal{S}^{Q}_{xy}(g) V_{xy}(g) \ket{\psi}\\
    &= V_{xy}^{\dagger}(g) \mathcal{S}^{Q}_{xy}(g) V_{xy}(g) V_{xy}(g^{-1}) U^{Q1}_{xy}(g) \ket{\psi}\\
    &= \omega_{xy}(g,g^{-1}) V_{xy}^{\dagger}(g) \mathcal{S}^{Q}_{xy}(g) [U^{ Q1}_{xy}(g^{-1})]^{\dagger} \ket{\psi}\\
    &= \omega_{xy}(g,g^{-1}) V_{xy}^{\dagger}(g) U^{Q2}_{xy}(g) \ket{\psi}\\
    &= V_{xy}(g^{-1}) U^{Q2}_{xy}(g) \ket{\psi} =  \ket{\psi}
\end{split}\label{eq:sqcomm}
\end{align}
where we have used $V_{xy}(g)V_{xy}(g^{-1}) = \omega_{xy}(g,g^{-1})$, and $V_{xy}^{\dagger}(g) = \omega^{*}_{xy}(g,g^{-1}) V_{xy}(g^{-1})$.
We remark that statements such as Eq.~\ref{eq:dangerous} are dangerous, as they deal with operators of infinite support acting on $\ket{\psi}$ (for which the overall phase factor is not-so-well-defined).  
Instead of using these infinite operators, we may instead replace them with finite rectangular operators with appropriately dressed corners,
\begin{align}
    \widetilde{U}_{xy}^{Q1}(g) &= V_{x_0 y}(g)V_{x_0 y_1}(g^{-1})V_{x y_1}(g) {U}_{x_0 x}^{y y_1}(g)\\
    \widetilde{U}_{xy}^{Q2}(g) &= V_{x y_0}(g)V_{x_1 y_0}(g^{-1}) V_{x_1 y}(g) {U}_{x x_1}^{y_0 y}(g)
\end{align}
for some $x_0\ll x \ll x_1$ and $y_0\ll y \ll y_1$.
These satisfy Eq.~\ref{eq:dangerous} exactly, and $\widetilde{\mathcal{S}}^{Q}_{xy}(g)\equiv \widetilde{U}_{xy}^{Q1}(g^{-1})\widetilde{U}_{xy}^{Q2}(g)$ acts in the same way as $\mathcal{S}^{Q}_{xy}(g)$ near $V_{xy}(g)$.
The important fact is that these operators only touch at $(x,y)$, so the other corners may effectively be ignored and we arrive at the same result.
From this, we conclude that $S_{xy}^{Q}(g)$ commutes with $V_{xy}(g)$ when acting on $\ket{\psi}$.  

Finally, since $\mathcal{S}^{Q}_{xy}(g) = \mathcal{S}^{R}_x(g) [\mathcal{S}^{ T}_y(g)]^{\dagger}$, 
we have from the definition of $\beta^{R/T}_{xy}$ that
\begin{equation}
\beta_{xy}^{R}(g)\beta_{xy}^{*T}(g) = 1
\end{equation}
With all these parts combined, we may conclude that $\beta(g) = \beta_{xy}^{R}(g) = \beta_{xy}^{T}(g)$ does not depend on $x$, $y$, or $T$/$R$.
Analogous arguments also show that
$\beta_{xy}^{L}(g) = \beta_{xy}^{B} (g) = \beta(g)$.

It then follows that $\beta(g)$ cannot be changed by an LSLU evolution.  
A local symmetric unitary cannot transform $\beta(g)$ throughout the entire system at once, for the same reason it could not change $\phi$ for a 1D SPT.
Similarly, a linearly-symmetric local unitary may make changes to quantities defined along whole lines but cannot make a global change that would affect $\beta(g)$.

We remark that such a result does not hold for other similar quantities.  For example, the phase obtained from commuting a single line symmetry, $S^{v}_{x^\prime}(g^\prime)$, with $V_{xy}(g)$, may be non-trivial if $x^\prime$ is near $x$.  
This phase is independent of $y$ and therefore cannot be changed by a SLU evolution.  However, it \emph{can} be changed by an LSLU evolution, which acts along the entire column at once.
Also, the phase obtained from commuting $\mathcal{S}^{R}_x(g^\prime)$ with $V_{xy}(g)$, for $g^\prime \neq g$, need not be the same as for $\mathcal{S}^{T}_y(g^\prime)$.  
This is therefore again only a property of a line, and can be changed by an LSLU evolution.
Only those phases $\beta(g)$ coming from $g^\prime = g$ are bulk properties and therefore conserved under LSLU evolution.
Note that this procedure is isomorphic to observing the charge response of the symmetry $\mathcal{S}^{R}_x(g)$ to a \emph{twist} of the symmetry $\mathcal{S}^{L}_x(g)$~\cite{Levin2012-dv,Wen2017-ak}.

A question still remains as to what consistent choices are possible for $\beta(g)$.
This will lead to a classification of all strong equivalence classes of SSPT phases with onsite symmetry $G_s$.

\subsection{At the edge}\label{ssec:edge}
At this point, it is convenient to introduce an edge into our system at $y=1$ and $y=\ell_y$.  
This allows us to present an alternate view of our findings in the previous section.
We proceed to derive some of the same results, but from a different perspective.

After introducing edges, the ground state manifold becomes massively degenerate, with degeneracy growing exponentially as $\exp({\mathcal{O}(L_\mathrm{edge})})$, where $L_\mathrm{edge}$ is length of the edge~\cite{You2018-em}.
Similar to the case of 1D SPTs, a vertical subsystem symmetry may be decomposed into two operations acting on the top/bottom edge of the system,
\begin{equation}
    S^{v}_{x}(g) \ket{\psi} = V^\mathrm{top}_x(g) V^\mathrm{bot}_x(g) \ket{\psi}
\end{equation}
which operate within the ground state manifold.  
We focus on the group of vertical symmetries, an extensively large group $G_v=(G_s)^{L_\mathrm{edge}}$, with a linear representation generated by $\{S^v_x(g)\}$. 
In analogy to 1D, the representation of the symmetry group $G_v$ on the top edge, $V^\mathrm{top}_y(g)$, may be a projective representation.
Note that, unlike for 2D SPTs under global symmetries, the symmetries act locally at the edge and do not give rise to non-trivial 3-cocycles. 

Let $h_x^{g}\in G_v$ be the group element represented by $S^{v}_x(g)$, $\omega_\mathrm{top}(h,h^\prime)$ be the factor system of $V^\mathrm{top}_x(g)$, and define $\phi_\mathrm{top}(h,h^\prime) = \omega_\mathrm{top}(h,h^\prime)/\omega_\mathrm{top}(h^\prime,h)$ the phase obtained from commuting $h$, $h^\prime$. 
As the Hamiltonian is local, we may assume that $\omega_\mathrm{top}$ is a \emph{local} projective representation, which we define to be one such that $\phi_\mathrm{top}(h_x^g, h_{x^\prime}^{g^\prime})=1$ if $x$ and $x^\prime$ are separated by a distance much larger than the correlation length.
Equivalently, this means they can be brought into a form where $\omega_\mathrm{top}(h_x^g, h_{x^\prime}^{g^\prime})=1$ for far separated $x$, $x^\prime$.

Under LSLU evolution, the class of this projective representation may be changed ``locally'', subject to certain extra constraints.
By a ``local'' change in projective representation, we mean modifications to $\omega_\mathrm{top}$ that can be made up of consecutive single local changes,
where a single local change is one in which $\omega_\mathrm{top} \rightarrow  \omega_\mathrm{top} \omega_{loc}$ for 
$\omega_{loc}$ satisfying 
\begin{align}
\begin{split}
\phi_{loc}(h_x^{g}, h_{x^\prime}^{g^\prime}) &\equiv \frac{{\omega}_{loc}(h_x^{g}, h_{x^\prime}^{g^\prime})}{\omega_{loc}( h_{x^\prime}^{g^\prime},h_x^{g})}=1 \\
&\hspace{0.5cm} \text{if } x\notin [x_0,x_1] \text{ or } x^\prime \notin [x_0,x_1]
\end{split}\label{eq:philoc}
\end{align}
for some finite range $[x_0, x_1]$ on the order of the correlation length.
Note that a single local change is accomplished by a long vertical 1D unitary evolution that only respects the symmetries as a whole (a $U_{ls}$ from earlier).
Any change that can be made up of consecutive local changes is itself a local change, which can be implemented by an LSLU evolution. 
However, as alluded to earlier, there are some extra constraints that $\omega_\mathrm{top}$ must satisfy, which arise due to the requirement that the orthogonal horizontal symmetries $\{S^h_y(g)\}$ must also be respected.
Thus, we are interested in the equivalence class of local projective representations satisfying these constraints, modulo local changes.

The extra constraints may be thought of as the following: $V^\mathrm{top}_{x}(g)$ must commute with all $S^h_y(g^\prime)$, since the overall representation $V^\mathrm{top}_{x}(g)V^\mathrm{bot}_{x}(g)$ must be linear, and the horizontal symmetries may only overlap with one of them at most.
At the same time, we have the identity
\begin{equation}
     \prod_{y=1}^{\ell_y} S^h_y(g^\prime)
    = \prod_{x=-\infty}^{\infty}  S^v_x(g^\prime)
\end{equation}
which implies that
\begin{equation}
    \left[ V^\mathrm{top}_x(g), \prod_{x^\prime=-\infty}^{\infty} V^\mathrm{top}_{x^\prime}(g^\prime)\right]\ket{\psi}=0
\end{equation}
placing a constraint on possible classes of projective representations $\omega_\mathrm{top}$.
In terms of $\phi_\mathrm{top}$, this implies 
\begin{equation}
\prod_{x^\prime=-\infty}^{\infty} \phi_\mathrm{top}(h^{g}_x,h^{g^\prime}_{x^\prime}) = 1 .\label{eq:phitconstraint}
\end{equation}
We remark that there are no issues with the $\infty$, as the representation is local and we may simply restrict the product over $x^\prime$ to some finite range about $x$.

All single local changes $\omega_{loc}, \phi_{loc}$, must also satisfy this constraint.
Take $\phi_{loc}$ to be non-trivial only within the range $[x_0,x_1]$, and let $x_\frac{1}{2}$ lie within this interval.
Then, observe the phase resulting from commuting ${h^g_\mathrm{left} = \prod_{x<x_\frac{1}{2}} h^{g}_x}$ with ${h^{g'}_\mathrm{right} = \prod_{x\geq x_\frac{1}{2}} h^{g^\prime}_x}$,
\begin{equation}
    \phi_{loc}(h^g_\mathrm{left}, h^{g^\prime}_\mathrm{right}) = \prod_{x<x_\frac{1}{2}}\prod_{x^\prime \geq x_\frac{1}{2}} \phi_{loc}(h^g_{x}, h^{g^\prime}_{x^\prime})
\end{equation}
using the fact that $\phi_{loc}$ must satisfy the same constraints as $\phi_\mathrm{top}$, multiplying by the conjugate of Eq.~\ref{eq:phitconstraint} we get
\begin{equation}
    \phi_{loc}(h^g_\mathrm{left}, h^{g^\prime}_\mathrm{right}) = \prod_{x_0\leq x<x_\frac{1}{2}}\prod_{x_0\leq x^\prime < x_\frac{1}{2}} \phi_{loc}^{*}(h^g_{x}, h^{g^\prime}_{x^\prime})
\end{equation}
Since $\phi_{loc}$ is only non-trivial with $[x_0,x_1]$, we have explicitly restricted $x$ and $x^\prime$ to this interval.
In the case where $g=g^\prime$, $\phi_{loc}(h^g_x,h^g_{x^\prime}) = \phi_{loc}^{*}(h^g_{x^\prime}, h^g_x)$, and so
\begin{equation}
    \phi_{loc}(h^g_\mathrm{left}, h^{g}_\mathrm{right}) = 1
\end{equation}
Hence, a local modification $\phi_\mathrm{top} \rightarrow \phi_\mathrm{top} \phi_{loc}$ cannot changed the value of $\phi_\mathrm{top}(h^g_\mathrm{left}, h^{g}_\mathrm{right})$.

It is possible that this value will be non-trivial in $\phi_\mathrm{top}$.
Consider putting periodic boundary conditions along the $x$ direction, identifying $x=0$ and $x={\ell_x}$, such that the overall topology is a cylinder.
Let $x_\frac{1}{2}$ be, say, near $\ell_x/2$.
If we define $h^{g}_\mathrm{left}$ and $h^{g}_\mathrm{right}$ to be products from $0$ to $x_\frac{1}{2}$ and from $x_\frac{1}{2}$ to $\ell_x$, respectively, we would similarly find that  $\phi_\mathrm{top}(h^g_\mathrm{left}, h^{g}_\mathrm{right}) = 1$.
However, as $\phi_\mathrm{top}$ is local, we may decompose $\phi_\mathrm{top}(h^g_\mathrm{left}, h^{g}_\mathrm{right})$ into a contribution coming from near $x_\frac{1}{2}$ and coming from the boundaries $0$ and $\ell_x$.  
To isolate the contribution coming from $x_\frac{1}{2}$, let us define $\xi\ll \ell_x $ to be some length for which $\phi_\mathrm{top}(h_x^{g},h_{x^\prime}^{g})$ is trivial if $|x-x^\prime|> \xi$.  
Then, redefining 
\begin{equation}
    \tilde{h}^{g}_\mathrm{left} = \prod_{(x_\frac{1}{2}-\xi) \leq x < x_\frac{1}{2}} h^{g}_x, \hspace{0.5cm}
    \tilde{h}^{g}_\mathrm{right} = \prod_{x_\frac{1}{2} \leq x < (x_\frac{1}{2}+\xi)} h^{g}_x
\end{equation}
we find that $\phi_\mathrm{top}(\tilde{h}^g_\mathrm{left}, \tilde{h}^{g}_\mathrm{right})$ need not be 1.
In fact, 
\begin{equation}
     \phi_\mathrm{top}(\tilde{h}^g_\mathrm{left}, \tilde{h}^{g}_\mathrm{top}) = \beta(g)
\end{equation}
is exactly our bulk invariant from earlier.
This can be seen (similar to in 1D) by placing a side of the truncated symmetry operator $U_{x_0 x_1}^{y_0 y_1}(g)$ along an edge.
It then follows from our previous proof that this phase is independent of where the cut $x_\frac{1}{2}$ is made. Furthermore, the phase is insensitive to the orientation of the cylinder and cut. 

\subsection{Classification}
\label{subsec:classification}
Let us now discuss the possible consistent choices for $\beta(g)$, and in this way classify all strong equivalence classes of SSPT phases.

In the previous section we reduced the 2D bulk physics down to the 1D problem of local projective representations along an edge, and finally down to a 0D problem involving $h^{g}_\mathrm{left/right}$ about a single cut in the edge.
In this final picture, we are essentially examining properties of the projective representation $\omega_\mathrm{top}$, $\phi_\mathrm{top}$,  of the group $G_s^2 = G_s^\mathrm{left}\times G_s^\mathrm{right}$.
Certain parts of this representation, namely $\beta(g) = \phi_\mathrm{top}(h^{g}_\mathrm{left},h^{g}_\mathrm{right})$, are universal throughout the system and invariant under LSLU transformations, and hence define the equivalence class.
The different equivalence classes are therefore in one-to-one correspondence with \emph{projective representations of $G_s^{2}$, modulo changes that leave $\beta(g)$ invariant}.

Let us denote by the superscript $g^{L(R)}$ the element $g$ from $G_s^\mathrm{left(right)}$, and $\omega$ a factor system of $ G_s^\mathrm{left}\times G_s^\mathrm{right}$.  
Consider modifying $\omega\rightarrow \omega \tilde{\omega}$.
There are three classes of $\tilde{\omega}$ that leave $\beta(g)=\omega(g^L,g^R)/\omega(g^R,g^L)$ unchanged.
Let $\omega_0$ be a factor system for any projective representation of $G_s$,
\begin{enumerate}
    \item  We may define $\tilde{\omega}(g^{L}_1 g^{R}_2, g^{L}_3 g^{R}_4) = \omega_0(g_1,g_3)$.
    This trivially leaves $\beta(g)$ unchanged, as 
    $\tilde{\omega}(g^L,g^R) =  \tilde{\omega}(g^R,g^L) =  1$.
    \item We may also use
    $\tilde{\omega}(g^{L}_1 g^{R}_2, g^{L}_3 g^{R}_4) = \omega_0(g_2,g_4)$.
    \item Finally, we may again use $\omega_0$ to define
    \begin{equation}
    \tilde{\omega}(g^{L}_1 g^{R}_2, g^{L}_3 g^{R}_4) = \frac{\omega_0(g_1 g_2, g_3 g_4)}{\omega_0(g_1, g_3)\omega_0(g_2, g_4)}\label{eq:omegatilde}
    \end{equation}
    This is independent of the previous two, satisfies the factor system condition (Eq.~\ref{eq:factorsystem}),
    and leaves $\beta(g)$ invariant, as 
    $\tilde{\omega}(g^L,g^R) = \tilde{\omega}(g^R,g^L) = \omega_0(g,g)$.
\end{enumerate}
Therefore, we want the projective representations of $G_s^2$, classified as $\mathcal{H}^2[G_s^2,U(1)]$,
modulo these three types of transformations, each of which are classified according to $\mathcal{H}^2[G_s,U(1)]$.
This is shown graphically for the explicit example of $G_s=\mathbb{Z}_2\times\mathbb{Z}_2$ in the next section. We remark that Eq.~\eqref{eq:omegatilde} is unambiguous since Eq.~\eqref{eq:symdef} specifies an isomorphism between any pair of onsite groups $G_s$. 

This leads us to the second main result in this paper:
\emph{The possible choices of $\beta(g)$, and therefore distinct strong equivalence classes of SSPT phases, are in one-to-one correspondence with elements of the group 
\begin{equation}
\mathcal{C}[G_s] = \mathcal{H}^2[G_s^2,U(1)]/(\mathcal{H}^2[G_s,U(1)])^3.
\label{eq:classification}
\end{equation} } 

The group structure is induced by a stacking operation.  
Consider two strong SSPT phases with onsite symmetry $G_s$, characterized by $\beta_1(g)$, $\beta_2(g)$, corresponding to two elements $c_1,c_2\in\mathcal{C}[G_s]$.
Let us stack these two SSPTs, such that the local Hilbert space at site $(x,y)$ is $\mathcal{H}_{xy} = \mathcal{H}_{xy,1}\otimes\mathcal{H}_{xy,2}$ and the onsite symmetry acts as $u_{xy}(g) = u_{xy,1}(g)\otimes u_{xy,1}(g)$.
The number of rows or columns, and therefore the number of symmetries, is unchanged in this process.
For the resulting stacked system, $\beta(g) = \beta_1(g) \beta_2(g)$, which corresponds to the element $c=c_1c_2$ following the group structure of $\mathcal{C}[G_s]$.

We note that there is an alternate (perhaps more intuitive) formulation~\footnote{This was pointed out by a referee, which we are thankful for.} of the classification $\mathcal{C}[G_s]$.
Let $\mathcal{A}[G]$ be the Abelian group of all bilinear functions $G \times G \rightarrow U(1)$, meaning functions  satisfying $a(g_1 g_2, g_3) = a(g_1,g_3) a(g_2,g_3)$ and $a(g_1,g_2 g_3) = a(g_1,g_2) a(g_1,g_3)$, for $g_i\in G$.
Then, let $\mathcal{A}^{\mathrm{anti}}[G]$ be the subgroup of $\mathcal{A}[G]$ consisting of functions $a$ which satisfy $a(g,g)=1$ (or, equivalently, $a(g_1, g_2) = a(g_2,g_1)^{-1}$).
Then, the classification is given by $\mathcal{C}[G_s] = \mathcal{A}[G_s]/\mathcal{A}^{\mathrm{anti}}[G_s]$, which one can verify is equivalent to Eq~\ref{eq:classification}.

Actually computing $\mathcal{C}[G_s]$ for a particular group $G_s$ is straightforward, and done in Appendix~\ref{app:comp}.

\section{Example: 2D cluster model}\label{sec:examples}
\begin{figure}
    \centering
\includegraphics[width=0.4\textwidth]{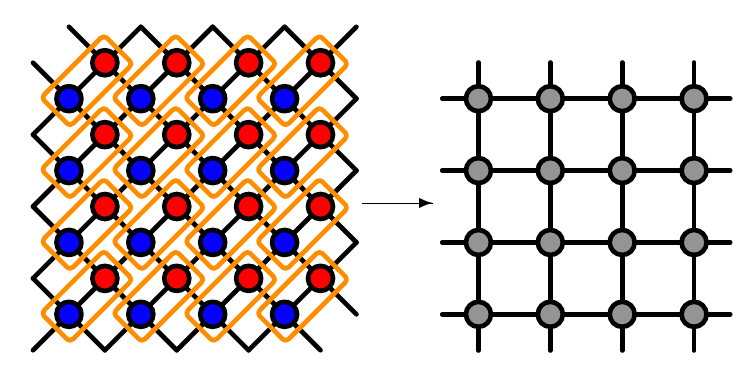}
    \caption{Illustration of the rotated square lattice on which the cluster model is originally defined.
    In order to bring the symmetries into the form considered here (Eq.~\ref{eq:symdef}), we combine two qubits into a single site, which we label as $a$ (blue) and $b$ (red), forming the square lattice shown on the right.
    }
    \label{fig:2dclus}
\end{figure}

Our statements thus far have been quite general.  
Let us now focus on the canonical example of an  SSPT~\cite{You2018-em}, the 2D cluster model on the square lattice.

While the symmetries of this model act along the diagonals of the original square lattice, we can easily bring them into the form considered here by rotating $45^\circ$ and forming a unit cell of two qubits, which we label by $a$ and $b$.
The Hamiltonian is given by
\begin{align}
H_\mathrm{clus} &= -\sum_{xy} X_{xy}^{(b)} Z_{xy}^{(a)} Z_{x+1,y}^{(a)} Z_{x,y+1}^{(a)} Z_{x+1,y+1}^{(a)}\\
&-\sum_{xy} X_{xy}^{(a)} Z_{xy}^{(b)} Z_{x-1,y}^{(b)} Z_{x,y-1}^{(b)} Z_{x-1,y-1}^{(b)}
\end{align}
where $X_{xy}^{(\alpha)}$ is the $X$ Pauli matrix acting on the $\alpha$ spin on site $(x,y)$, and similarly for $Z$.
This model is composed of mutually commuting terms, and thus can be solved exactly.

The onsite symmetry group $G_s$ is 
\begin{equation}
G_s=\mathbb{Z}_2\times\mathbb{Z}_2 = \{1,g_a,g_b,g_a g_b\}
\end{equation}
where $g_a$, $g_b$, are defined as the two generators of $G_s$.
The qubit degrees of freedom transform under this symmetry as
\begin{equation}
u_{xy}(g_a) = X_{xy}^{(a)}, \
u_{xy}(g_b) = X_{xy}^{(b)}
\end{equation}
and therefore lead to the generators of our subsystem symmetries
\begin{align}
S^{v}_{x}(g_\alpha) &= \prod_{y=-\infty}^{\infty} X_{x y}^{(\alpha)}, \
S^{h}_{y}(g_\alpha) = \prod_{x=-\infty}^{\infty} X_{x y}^{(\alpha)}\label{eq:clussyms}
\end{align}
for $\alpha\in\{a,b\}$,
which one can readily verify all commute with $H_\text{clus}$.
Each row/column is therefore associated with two symmetry generators.

Suppose we take the system on a torus of dimensions $L_x\times L_y$, in which case
there are $2(L_x+L_y)$ symmetry generators from Eq.~\ref{eq:clussyms}.
However, not all symmetries are unique, as we have
\begin{align}
\prod_{x_0=1}^{L_x} S^v_{x_0}(g_{\alpha}) = \prod_{y_0=1}^{L_y} S^h_{y_0}(g_\alpha) = \prod_{xy} X_{xy}^{(\alpha)} 
\end{align}
for each $\alpha\in\{a,b\}$.
The total symmetry group is therefore only $G = (\mathbb{Z}_2\times\mathbb{Z}_2)^{2(L_x+L_y-1)}$.

\begin{figure}
    \centering
    \definecolor{col1}{RGB}{90,108,255}
    \includegraphics[width=0.3\textwidth]{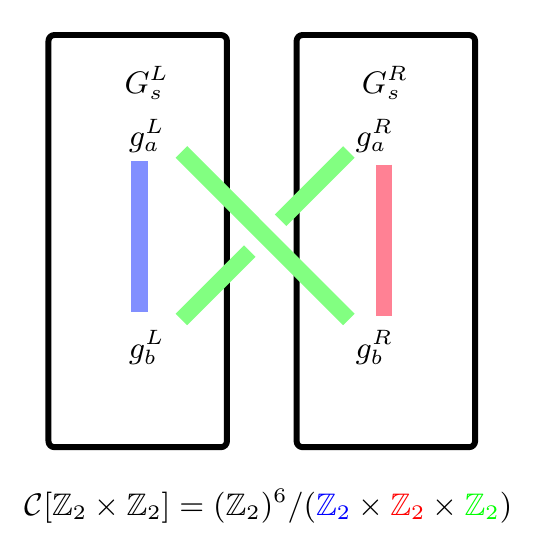}
    \caption{A graphical representation of the strong classification of the $G_s=\mathbb{Z}_2\times\mathbb{Z}_2$ model.
    The four generators of $G_s^2 = G_s^\mathrm{left}\times G_s^\mathrm{right}$ are indicated by $g_a^L,g_b^L,g_a^R,g_b^R$.
    Following Eq.~\ref{eq:classification}, we first compute $\mathcal{H}^2[G_s^2,U(1)] = (\mathbb{Z}_2)^6$, corresponding to a freedom to choose $\phi(g,g^\prime)=\phi(g^\prime,g)=\pm1$ for each pair of generators.
    The LSLU invariant quantities are $\beta(g)=\phi(g^L,g^R)$ between the same element $g$ from $G_s^L$ and from $G_s^R$ (for any $g\in G_s$, not just generators).
    The three types of non-trivial changes that leave all $\beta(g)$ invariant (given by near Eq.~\ref{eq:omegatilde}) act on these generator $\phi(g,g^\prime)$ as $(1)$ $\phi(g_a^L,g_b^L)\rightarrow -\phi(g_a^L,g_b^L)$, $(2)$ $\phi(g_a^R,g_b^R)\rightarrow -\phi(g_a^R,g_b^R)$, or $(3)$ $\phi(g_a^L,g_b^R),\phi(g_b^L,g_a^R)\rightarrow -\phi(g_a^L,g_b^R),-\phi(g_b^L,g_a^R)$ (illustrated by blue, red, and green).
    The overall classification is therefore obtained by modding out three copies of $\mathcal{H}^2[G_s,U(1)]=\mathbb{Z}_2$ from $(\mathbb{Z}_2)^6$, resulting in $\mathcal{C}[(\mathbb{Z}_2)^2] = (\mathbb{Z}_2)^3$.
    }
    \label{fig:z2z2}
\end{figure}

Let us first probe the nontriviality of this phase in the bulk according to the procedure in Sec.~\ref{ssec:bulk}.
Construct the rectangular truncated symmetry operator,
\begin{equation}
    U_{x_0x_1}^{y_0y_1}(g) = \prod_{x=x_0}^{x_1-1}\prod_{y=y_0}^{y_1-1}u_{xy}(g)
\end{equation}
which creates excitations at the corners.
These excitations may be locally annihilated by operators $V_{xy}(g)$ at the bottom left and top right corners, and $V_{xy}(g^{-1})$ on the remaining two, given by
\begin{align}
    &V_{xy}(g_a) = Z^{(b)}_{x-1,y-1}, \hspace{0.5cm} V_{xy}(g_b) = Z^{(a)}_{xy} \\
    &V_{xy}(g_a g_b) = Z^{(b)}_{x-1,y-1}Z^{(a)}_{xy} 
\end{align}
and in the case of $\mathbb{Z}_2$, $g = g^{-1}$.
Note that there is some freedom in choosing $V$, and we have made a choice in this definition.
Calculating the invariants $\beta(g)$ (which are independent of our choice of $V$) using Eq.~\ref{eq:betadef}, we find
\begin{align}
    \beta(1) = \beta(g_a) = \beta(g_b) = 1,\hspace{0.5cm}
    \beta(g_a g_b) = -1
\end{align}
Since $\beta(g_ag_b)\neq 1$, this phase is indeed a non-trivial strong SSPT.
Utilizing Eq.~\ref{eq:classification}, the classification of strong SSPTs with this symmetry group is given by
\begin{equation}
    \mathcal{C}[\mathbb{Z}_2\times\mathbb{Z}_2] = \mathbb{Z}_2\times\mathbb{Z}_2\times\mathbb{Z}_2
    \, .
\end{equation}
This calculation may be understood graphically as described in Figure~\ref{fig:z2z2}.
In this case, each of $\beta(g_a)$, $\beta(g_b)$, and $\beta(g_a g_b)$ may be chosen independently, giving rise to a total of eight possible equivalence classes.

We may also arrive at this conclusion by examining the edge.
Consider a top edge at $y=\ell_y$.  
A symmetry $S^v_x(g)$ acts on the top edge as
\begin{align}
    V^\mathrm{top}_x(g_a) &= Z^{(b)}_{x-1,\ell_y} Z^{(b)}_{x,\ell_y} \\
    V^\mathrm{top}_x(g_b) &= Z^{(a)}_{x,\ell_y} X^{(b)}_{x,\ell_y} Z^{(a)}_{x+1,\ell_y}\\
    V^\mathrm{top}_x(g_ag_b) &= Z^{(b)}_{x-1,\ell_y} Z^{(a)}_{x,\ell_y} Z^{(b)}_{x,\ell_y} X^{(b)}_{x,\ell_y} Z^{(a)}_{x+1,\ell_y}
\end{align}
which forms a projective representation for large total vertical symmetry group $G_v = G_s^{L_\mathrm{edge}}$.
As before, let $h_x^g \in G_v$ be the element represented by $V^\mathrm{top}_x(g)$.  
This projective representation is characterized by the non-commuting pairs of generators, i.e. those with the non-trivial $\phi_\mathrm{top}$s,
\begin{equation}
    \phi_\mathrm{top}(h_x^{g_a}, h_{x}^{g_b}) =
    \phi_\mathrm{top}(h_x^{g_b}, h_{x+1}^{g_b}) = -1
\end{equation}
for all $x$, and the property  $\phi_\mathrm{top}(h,h^\prime)=\phi_\mathrm{top}(h^\prime,h)$.  
Notice that the set of $\phi_\mathrm{top}(h,h^\prime)$ between all generators of $G_v$ provides a complete description of the projective representation, as (see Appendix of Ref.~\onlinecite{You2018-em} for details) the elements of
\begin{equation}
    \mathcal{H}^2[\mathbb{Z}_2^{n},U(1)] = \mathbb{Z}_2^{\frac{n^2-n}{2}}
\end{equation}
are in one-to-one correspondence with different choices of $\phi_\mathrm{top}(h,h^\prime)$ for the $n$ generators of $\mathbb{Z}_2^n$, where $n=2L_\mathrm{edge}$.

\begin{figure}
    \centering
\includegraphics[width=0.5\textwidth]{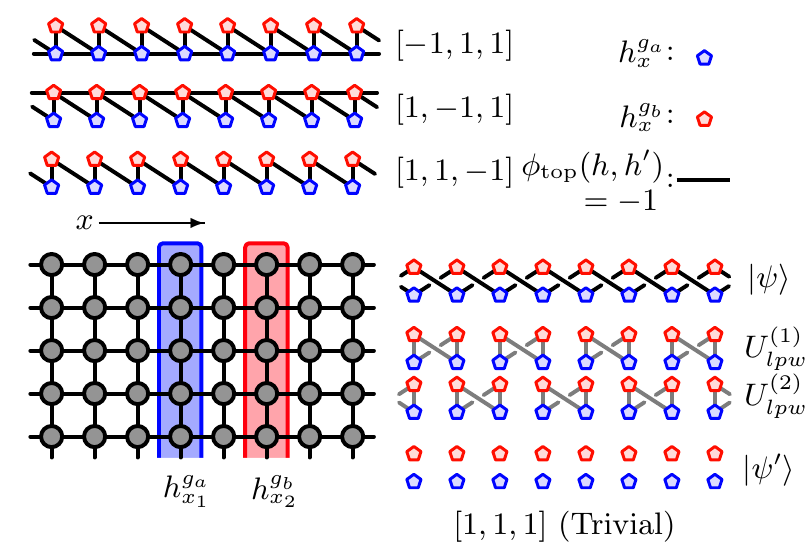}
    \caption{A graphical representation of a particular state may be obtained by looking at the projective representation of vertical symmetries terminating at the top edge (bottom left).
    The examples here are for the square lattice cluster model $G_s = \mathbb{Z}_2\times\mathbb{Z}_2$.
    (top) The generators of $G_v = G_s^{L_\mathrm{edge}}$, $h_x^{g_a}$ and $h_x^{g_b}$, are denoted by the blue and red markers respectively, and ordered by increasing $x$.  
    A link is drawn between generators if they have non-trivial commutation relations, $\phi_\mathrm{top}(h,h^\prime)=-1$.  
    The graphical representation of three representative states from distinct non-trivial equivalence classes are shown, labeled by $[\beta(g_a),\beta(g_b),\beta(g_ag_b)]$.
    Note that the representative states here have a translation-invariant projective representation --- this need not be the case for a general state.
    (bottom right) We show a state, $\ket{\psi}$, with a non-trivial projective representation on the edge, that nevertheless belongs to the weak (trivial) $[1,1,1]$ equivalence class, and how it may be transformed into the trivial projective representation via an LSLU composed of two linearly piecewise symmetric unitaries $U_{lpw}^{(1)}$ and $U_{lpw}^{(2)}$.
    The graphical representation shown for $U_{lpw}^{(i)}$ is that of the state $U_{lpw}^{(i)}\ket{+}$, where $\ket{+}$ is the trivial symmetric state (see Sec~\ref{sec:stacking}).
    }
    \label{fig:edgegraph}
\end{figure}
\begin{figure}
    \centering
\includegraphics[width=0.4\textwidth]{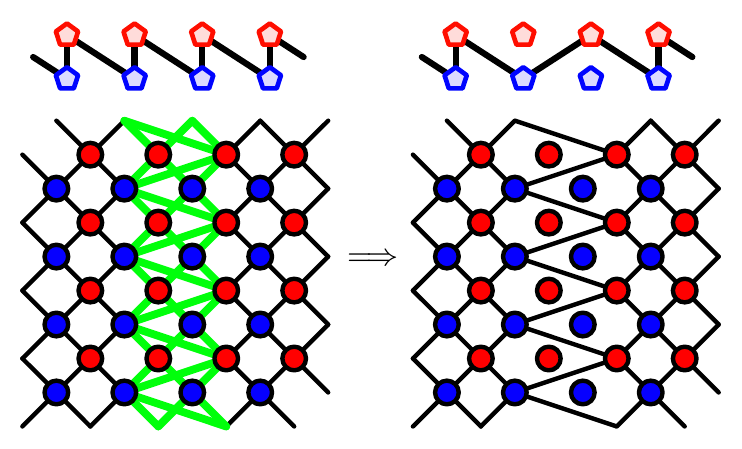}
    \caption{
    An explicit example of an LSLU composed of a single  $U_{ls} = \prod_{\langle i j\rangle} CZ_{ij}$, acting on the square lattice cluster model.
    The product is over all green links $\langle i j \rangle$ connecting qubits $i$ and $j$ and $CZ_{ij}$ is the controlled-Z gate acting on the two qubits.
    One can verify that $U_{ls}$ commutes with all subsystem symmetries (except at edges) as a whole, but cannot be written as a product of individually symmetric local gates (otherwise it would also be an SLU).  
    The resulting state after applying $U_{ls}$ can be thought of as a cluster state on a modified lattice, shown on the right.
    As a result of $U_{lpw}$, two columns have effectively been disentangled from the rest of the system.
    However, because the square lattice cluster state represents a non-trivial strong SSPT phase, the two subsystems on the left and right of $U_{lpw}$ are still entangled.  
    Above, we show the graphical picture of the projective representation (see Fig~\ref{fig:edgegraph} for legend).
    One can verify that before and after, $[\beta(g_a),\beta(g_b),\beta(g_a g_b)]=[1,1,-1]$ remains invariant.
    }
    \label{fig:lsluexample}
\end{figure}

A graphical understanding of the projective representation may be obtained by representing each generator, $h^{g_a}_x$ and $h^{g_b}_x$, as vertices along a line ordered by $x$. This is demonstrated in Fig.~\ref{fig:edgegraph}.
Two points $h$ and $h^\prime$ are connected by a link if $\phi_\mathrm{top}(h,h^\prime)=-1$.
The constraint (Eq.~\ref{eq:phitconstraint}) means that each point must be connected by a link to an even number of $h^{g_a}_x$, and an even number of $h^{g_b}_x$.  
The invariant $\beta(g_a)$ is obtained by cutting the line somewhere, and counting the parity of links between two $h^{g_a}$ vertices that were cut, and similarly for $\beta(g_b)$.  In this case, both are zero and so $\beta(g_a)=\beta(g_b)=1$.
The final invariant, $\beta(g_a g_b)$, is simply the total parity of links crossing the cut.  
In this case, there is one link, and so $\beta(g_a g_b)=-1$.

As an instructive example, consider the case of $G_s=\mathbb{Z}_N=\{1,g_a, g_a^2, \dots, g_a^{N-1}\}$.  
We draw a chain of $h^{g_a}_x$ ordered by $x$.
In this case, $\phi_\mathrm{top}(h, h^\prime)$ may be any $N$th roots of unity.
In the graphical representation, we may draw a directed link going from $h^{g_a}_x$ to $h^{g_a}_{x^\prime}$, and associate with each link a value $\phi(h^{g_a}_x, h^{g_a}_{x^\prime})$.
The constraint (Eq.~\ref{eq:phitconstraint}) demands that the total \emph{flux} going in to a vertex is equal to the outgoing flux mod $N$, i.e. the flux flow is \emph{divergenceless}.
The invariant $\beta(g_a)$ gives the net flux flow going along the length of the line.
Indeed, $\mathcal{C}[\mathbb{Z}_N]=\mathbb{Z}_N$, and there are $N$ total strong equivalence classes corresponding to the $N$ values the flux can take.
In this picture, it is immediately clear that $\beta(g_a)$ cannot be modified via a local change in the projective representation, as only a global action can change the total flux flow.

Going back to the cluster model with $G_s=\mathbb{Z}_2\times\mathbb{Z}_2$, we found that there were a total of eight strong equivalence classes.
What do the states in thes equivalence classes look like?  
We may represent an equivalence class by $[\beta(g_a),\beta(g_b),\beta(g_ag_b)]$.
The square lattice cluster state here corresponds to the $[1,1,-1]$ equivalence class.
The graphical representation for representative states from the other equivalence classes, $[-1,1,1]$ and $[1,-1,1]$, are shown in Figure~\ref{fig:edgegraph}.
All these strong equivalence classes may also be realized by commuting projector Hamiltonians.
For instance, the $[-1,1,1]$ equivalence class is realized by the Hamiltonian
\begin{align}
\begin{split}
H_{[-1,1,1]} &= -\sum_{xy} X_{xy}^{(b)} Z_{xy}^{(a)} Z_{x+1,y}^{(a)} Z_{x,y+1}^{(a)} Z_{x+1,y+1}^{(a)} \\
-\sum_{xy} &\left[X_{xy}^{(a)} Z_{xy}^{(b)} Z_{x-1,y}^{(b)} Z_{x,y-1}^{(b)} Z_{x-1,y-1}^{(b)}\right.\\
&\times \left.Z_{x+1,y+1}^{(a)} Z_{x+1,y-1}^{(a)} Z_{x-1,y+1}^{(a)} Z_{x-1,y-1}^{(a)}\right]
\end{split}
\end{align}

We show in Figure~\ref{fig:edgegraph}~(bottom right) an example of a state $\ket{\psi}$ in the trivial equivalence class $[1,1,1]$ that nevertheless realizes a seemingly non-trivial projective representation at the edge.  
This projective representation may be trivialized by two linearly piecewise unitaries $U_{lpw}$, which modify the $\phi_\mathrm{top}(h,h^\prime)$ of the state as shown.
The action of $U_{lpw}$ may be thought of as trivializing a stack of non-trivial  $(\mathbb{Z}_2)^{4}$ 1D SPT chains in the SSPT.
Thus $\ket{\psi}$ and a trivial state $\ket{\psi^\prime} = U_{lpw}^{(2)} U_{lpw}^{(1)} \ket{\psi}$ are connected via an LSLU evolution and therefore belong to the same (trivial) strong equivalence class.

Figure~\ref{fig:lsluexample} shows an explicit example of an LSLU acting on the square lattice cluster model.
The LSLU shown manages to disentangle two columns of qubits, but since the model represents a non-trivial strong SSPT, the two subsystems on the left and right of these columns are still entangled.  

Finally, a $G_s=\mathbb{Z}_n \times \mathbb{Z}_m$ generalization of the square lattice cluster model SSPT was introduced in Ref.~\onlinecite{You2018-em}.  
It was found that such a construction could give rise to $q\equiv\gcd(n,m)$ different phases (if $q=1$, the model was always trivial).
We now know that this model may be classified according to
\begin{equation}
    \mathcal{C}[\mathbb{Z}_n\times\mathbb{Z}_m] = \mathbb{Z}_n\times\mathbb{Z}_m\times\mathbb{Z}_{q}
\end{equation}
Each of the $q$ phases constructed in Ref.~\onlinecite{You2018-em} lie in distinct strong equivalence classes, and live within the final $\mathbb{Z}_q$ factor.
Thus, there are many more strong SSPT phases involving the $\mathbb{Z}_n$ or $\mathbb{Z}_m$ factor that were missed in the construction of Ref.~\onlinecite{You2018-em}.

\section{Other Aspects}\label{sec:aspects}
\subsection{Additional line-like subsystem}\label{sec:additional}
We may also consider systems with additional line-like subsystem symmetries.

For example, consider the cluster model on the triangular lattice.
We may redefine the unit vectors such that the triangular lattice is mapped on to the square lattice with additional connections going along the $\hat{x}+\hat{y}$ direction.
This Hamiltonian then takes the form
\begin{equation}
\begin{split}
    H_\mathrm{tri} = - \sum_{xy} &\left[X_{xy} Z_{x-1,y-1} Z_{x-1,y} Z_{x,y-1}\right. \\
     &\left. \times Z_{x,y+1} Z_{x+1,y} Z_{x+1,y+1}\right]
\end{split}
\end{equation}
which has the onsite symmetry group $G_s=\mathbb{Z}_2$, but now with three directions of subsystem symmetries: horizontal, vertical, and diagonal.
The diagonal symmetries are given by
\begin{align}
S^{d}_{q}(g) &= \prod_{x=-\infty}^{\infty} u_{x,x-q}(g)
\end{align}
where $q\in\mathbb{Z}$ corresponds to the different diagonals.

We must modify our definition of strong phase equivalence in this case.
It is natural to extend the definition of LSLU to allow for unitaries $U_{ls}$ along the diagonal $\hat{x}+\hat{y}$ direction.  
In general, we should allow for $U_{ls}$ to extend along any direction for which subsystem symmetries exist (in the case of line-like symmetries).

It is convenient to think in terms of edge projective representations.
Only symmetries going along the same direction may have non-trivial projective representations.
One can define a $\beta(g)$ from this projective representation, one for each of the three directions.  
It can then be shown like before that these three directions are not independent (and must be the same), and we are left with the same classification of $\beta(g)$ as before.
Thus, strong SSPTs with these extra subsystems also have a $\mathcal{C}[\mathbb{Z}_2] = \mathbb{Z}_2$ classification, and this model lies in the non-trivial phase.

\subsection{Adding or removing degrees of freedom}
In standard SPT phases protected by global symmetries, we are allowed to add or remove degrees of freedom. This is necessary to compare SPT phases on different system sizes. However, we are only allowed to add or remove degrees of freedom that transform as a linear representation of the symmetry.  
For example, the edge modes of the AKLT chain protected by time reversal symmetry can be gapped out if we add a spin-$1/2$ degree of freedom to the edges. 

In the case of SSPTs comparing phases on different system sizes is more subtle, as the total symmetry group increases with the system size. 
Consequently, it is neccesary to consider adding and removing degrees of freedom in several different ways.
We may locally add unentangled degrees of freedom to a site, as long as they transform linearly under $G_s$, which does not change anything.
However, we may also add an entire row or column at once, which actually increases the size of the total symmetry group.
To achieve this we may add an unentangled symmetric row of sites, for example, and each site should transform as a linear representation of $G_s$.
This defines a new horizontal symmetry acting on the new row, and existing vertical symmetries should be modified to act on this new row at their intersection.
Similarly, we can allow the removal of entire rows or columns, along with their symmetry, that are unentangled from the rest of the system. This allows us to meaningfully compare SSPTs on different lattice sizes, which lie in distinct conventional phases and have different total symmetry groups. Our strong equivalence relation successfully identifies SSPT models defined by the same local rule on different system sizes as belonging to the same equivalence class. 

\subsection{Blocking changes the symmetry structure}
With global SPTs, we may block multiple existing sites together to define a new site, without changing the structure of the symmetries.
We note here that the same is \emph{not} true for SSPTs.  
In particular, the choice of what defines a site is important.

For example, consider a model with a single spin on each site and a $G_s$ onsite symmetry group.
Suppose we take every odd row and combine it with the row below it, and combine each odd column with the one to its left.
Each unit cell now contains four spins.  
However, each row should now be associated with two types of horizontal symmetries, one which acts on the lower two qubits and another which acts on the upper two, and similarly for each column.
This does \emph{not} take the form of an SSPT as defined in Sec.~\ref{sec:setting}, even if we allow for a larger onsite symmetry group.

This motivates a more general definition of subsystem symmetries than discussed previously. 
For the above example, a natural generalization is to allow a triple of onsite abelian groups $G_\mathrm{both} \times G_\mathrm{horz} \times G_\mathrm{vert}$, where $G_\mathrm{both} \times G_\mathrm{horz}$ participate in horizontal symmetries and  $G_\mathrm{both} \times G_\mathrm{vert}$ participate in the vertical symmetries. 
In the above example $G_\mathrm{both}=G_s$ is given by the diagonal $G_s$ subgroup of the total onsite $(G_s)^4$ symmetry, of the form $(g,g,g,g)$ (labeling the onsite spins counterclockwise from the top right), where $g\in G_s$. 
Similarly, $G_\mathrm{horz}$ is given by another $G_s$ subgroup of the form $(g,g,1,1)$, and $G_\mathrm{vert}$ is given by the $G_s$ subgroup of the form $(1,g,g,1)$. 
As the LSLU equivalence relation does not care about choice of unit cell or specific symmetry structure,
such a coarse-graining cannot affect the overall classification which is therefore still given by $\mathcal{C}[G_s]$.
We claim that the classification for this generalized symmetry structure is given by simply $\mathcal{C}[G_\mathrm{both}]$, and is independent of $G_\mathrm{vert}$ or $G_\mathrm{horz}$.
Indeed, the projective representations at an edge involving $G_\mathrm{horz}$ or $G_\mathrm{vert}$ are not subject to as strong constraints as those placed on $G_\mathrm{both}$, and can always be trivialized via LSLU (specifically, the constraint in Eq.~\ref{eq:phitconstraint} only has to hold for $g^\prime\in G_\mathrm{both}\times G_\mathrm{vert}$ which have a non-trivial $G_\mathrm{both}$ component). 
This means, for example, that models with subsystem symmetries only along only one direction (e.g. $G_\mathrm{both}=G_\mathrm{horz}=\mathbb{Z}_1$, but $G_\mathrm{vert}$ is non-trivial) are always weak SSPTs.

In the above coarse-graining example we could instead chose to preserve only the $G_\mathrm{both} = G_s$ onsite symmetry group. This achieves a mapping between SSPTs on lattices of different scales with the same onsite group, but different representations. This highlights that symmetry-respecting real-space renormalization of SSPTs is a subtle issue which we plan to deal with in a forthcoming work~\cite{subsystemRG}.

\subsection{Equivalence of LSLU and stacking with weak SSPTs}\label{sec:stacking}
One useful perspective on the effect of allowing LSLUs, as opposed to simply SLUs, is that equivalence under LSLUs may be thought of as equivalence under a combination of SLUs and stacking with weak SSPTs.
Take the linearly supported symmetric unitary $U_{ls}$, which acts upon a horizontal or vertical line (which may encompass multiple rows or columns), and commutes with all symmetries as a whole.
Acting on the trivial symmetric product state, denoted by $\ket{+}$, $U_{ls}\ket{+}$ may describe a non-trivial 1D SPT state running along this line.
From the perspective of the symmetry action at the edge, there is no difference between acting with the unitary on the state, $\ket{\psi}\rightarrow U_{ls}\ket{\psi}$, versus stacking with this 1D SPT state, $\ket{\psi}\rightarrow\ket{\psi}\otimes U_{ls}\ket{+}$ (and extending the onsite representation appropriately).
Stacking with a disjoint set of such 1D SPT chains is then identical to a linearly piecewise unitary $U_{lpw}$, and allowing for multiple layers of such stacks captures the effect of an arbitrary LSLU evolution.
One should also allow for local unitaries and isometries that can reduce the local Hilbert space dimension in this picture.
Note that we have assumed here that any 1D SPT may be be created by a local unitary circuit acting on the product state --- this is true in 1D (but not in higher dimensions where other types of SPTs exist~\cite{Kapustin2014-tn}).

For example, consider the weak phase with a highly non-trivial projective representation at the edge, as in Fig.~\ref{fig:edgegraph}~(bottom right).  
The action of $U_{lpw}^{(1)}$ may be thought of as stacking with a weak SSPT (the fact that it is weak is clear from the disjointness of the edge projective representation), and similarly $U_{lpw}^{(2)}$.  
Stacking these two phases on top of the initial phase produces one with a trivial linear representation, which may then be brought to a trivial product state via SLUs (once the horizontal symmetries have similarly been brought to a linear representation at the edge).
For a strong SSPT phase, this is not possible.

This shows that the equivalence relation defined by LSLUs indeed coincides with the intuitive defition of a strong or weak SSPT.
We may define a \emph{disjoint} SSPT as a subclass of weak SSPTs, which is one such that the projective representation along the edge may be separated into those coming from disjoint sets of rows or columns.
This is the intuitive definition of a weak SSPT employed in Ref.~\onlinecite{You2018-em}.  
Stacking two disjoint SSPTs will generally result in a weak (but not necessarily disjoint) SSPT.
If one wishes for disjoint SSPTs to be weak, and for weak SSPTs to be closed under stacking, then one is lead to precisely the equivalence relation proposed in this paper. 


\subsection{Spurious topological entanglement entropy}\label{sec:spurious}
Recently, a connection was made~\cite{subsystemTEE} between SSPTs and spurious values of topological entanglement entropy~\cite{KitaevPreskill,levinwenentanglement} (TEE) found in the bulk of certain short-range entangled 2D phases~\cite{PhysRevB.94.075151}. Here we show that strong SSPT phases always lead to spurious values of TEE. 

\begin{figure}[t]
    \centering
\includegraphics[width=0.25\textwidth]{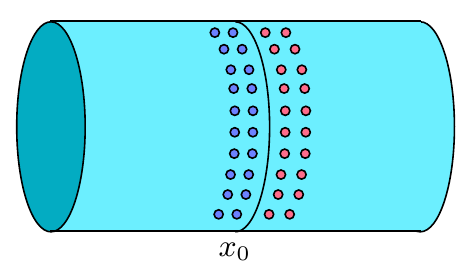}
    \caption{
    The cylindrical setup used in the cylinder extrapolation method to calculation the TEE.  
    The system is put on a cylinder, and the entanglement entropy across the cut dividing the system into two halves is calculated.  
    Via unitary evolution $U_{\mathrm{left}(\mathrm{right})}$ with support only on the left (right) half, all the sites far away from the cut may be disentangled into the trivial product state.  
    The remaining entangled sites are near the cut and shown as blue/red circles.  
    The calculation of entanglement entropy is therefore reduced to that of a 1D system but with an extensive bipartitioning into left/right partitions.
    }
    \label{fig:cylinder}
\end{figure}

One of the examples given in Ref.~\onlinecite{PhysRevB.94.075151} is that of the triangular lattice cluster state (which we have shown in Sec~\ref{sec:additional} belongs to a non-trivial strong SSPT phase protected by three directions of linear symmetries).
It was noted that using the cylinder extrapolation method~\cite{Jiang2012-oj} for this state leads to a spurious non-zero value of the TEE, despite the lack of topological order 
(Ref.~\onlinecite{PhysRevB.94.075151} only found spurious contributions via the cylinder extrapolation method, but they have since also been noted to occur for SSPTs via the general $A,B,C$ partitioning methods~\cite{KitaevPreskill,levinwenentanglement} when the boundaries of the chosen partitions run along the directions of subsystem symmetry~\cite{subsystemTEE}).
In the cylinder extrapolation method, the 2D system is taken on a cylinder of circumference $L$ along (say) the vertical direction, and bipartitioned into a left and right half as in Fig.~\ref{fig:cylinder}.  
The TEE $\gamma$ is obtained from the limit $\gamma = -S(L=0)$.
Ref.~\onlinecite{PhysRevB.94.075151} reduced the calculation of the entanglement entropy of the 2D system down to that of 1D system going along the cut, but with an extensive bipartitioning.
It was found that this 1D system exhibited an additional $\mathbb{Z}_2\times \mathbb{Z}_2$ symmetry, and was a non-trivial 1D SPT \emph{under the product group}.
A 1D SPT with symmetry $G = G_1\times G_2$ is defined to be non-trivial under the product group if there exists
\begin{equation}
    \phi(g,h)\equiv \frac{\omega(g,h)}{\omega(h,g)}\neq 1, \; \; \; g\in G_1, h\in G_2\label{eq:nontrivialprod}
\end{equation}
where $\omega$ is the factor system characterizing the 1D SPT phase.
It was shown generally that the 1D system appearing at the cut being non-trivial under the product group $G_1\times G_2$, where $G_1$ acts only on the left and $G_2$ only on the right of the cut, is a sufficient condition for a non-zero spurious TEE.
It is no coincidence that this is reminiscent of our strong classification, which relied on a particular non-trivial projective representation of the product group $G_s^\mathrm{left}\times G_s^\mathrm{right}$.

Quite generally, consider an SSPT state $\ket{\psi}$  with onsite symmetry group $G_s$, on a cylinder with circumference $L_y$ along the vertical ($y$) direction, and infinite extent in the horizontal ($x$) direction.
Consider calculating the entanglement entropy $S(\ket{\psi})$ across a cut at $x=x_0$ dividing the system into a left and right  half.
Let us define the  symmetry operations
\begin{equation}
    S^\mathrm{left}(g) = \prod_{x<x_0} S^v_x(g), \; \; \;
    S^\mathrm{right}(g) = \prod_{x\geq x_0} S^v_x(g)\label{eq:sleftright}
\end{equation}
for $g\in G_s$, which act on the left/right side of the cut.
Since the entanglement entropy is invariant under unitary operations localized on either side, $S(\ket{\psi})=S(U_\mathrm{left} U_\mathrm{right} \ket{\psi})$, we may simply choose $U_{\mathrm{left}(\mathrm{right})}$ such that the system becomes the trivial product state away from the cut (here $U_{\mathrm{left}(\mathrm{right})}$ is only supported on the $\mathrm{left}(\mathrm{right})$ half).
Let us also enforce that $U_{\mathrm{left}(\mathrm{right})}$ commute with all $S^{\mathrm{left}(\mathrm{right})}(g)$ \emph{as a whole} (note that this does not interfere with the ability to disentangle spins away from the cut, as they are being brought to the trivial symmetric product state).
This reduces the calculation down to that of a 1D system going along the cut, as shown in Fig.~\ref{fig:cylinder}, with an extensive bipartitioning.
As $U_{\mathrm{left}(\mathrm{right})}$ commutes with $S^{\mathrm{left}(\mathrm{right})}(g)$, this remaining 1D system is itself a 1D SPT and is symmetric under the symmetry group $G_s^\mathrm{left}\times G_s^\mathrm{right}$.
The first factor ($G_s^\mathrm{left}$) acts on the left (blue) sites, and the second ($G_s^\mathrm{right}$) acts on the right (red) sites in Fig.~\ref{fig:cylinder}.
The result in Ref.~\onlinecite{PhysRevB.94.075151} implies that if this 1D SPT is non-trivial \emph{under the product group} $G_s^\mathrm{left} \times G_s^\mathrm{right}$, then there will be a non-zero spurious contribution to the calculated TEE.

By our classification, an SSPT is strong if there exists 
\begin{equation}
    \beta(g) = \frac{\omega(g^L,g^R)}{\omega(g^R,g^L)}\neq 1,
\end{equation}
where $g^L$ and $g^R$ correspond to the same group element $g\in G_s$ from $G_s^\mathrm{left}$ and $G_s^\mathrm{right}$, respectively.
We therefore see immediately from Eq.~\ref{eq:nontrivialprod}, with $G_1=G_s^\mathrm{left}$ and $G_2=G_s^\mathrm{right}$ that a non-trivial $\beta(g)$ necessarily implies that the 1D SPT is non-trivial under the product group, which therefore implies a non-zero spurious contribution to the TEE.  That is,
\begin{equation}
    \textrm{Strong SSPT} \implies \textrm{Spurious TEE}\label{eq:sspttee}
\end{equation}
when measured using the cylinder extrapolation method, if the cut lies parallel to a subsystem direction.
However, the converse implication is not true: a non-zero spurious TEE in an SSPT does not imply that the SSPT is strong.
A zero TEE implies that the projective representation is \emph{disjoint} across the cut $x_0$ (there are no lines crossing the cut in the graphical representation), which implies that the SSPT is weak (this is simply the transposition of Eq.~\ref{eq:sspttee}).

\section{Conclusion}\label{sec:conclusion}
We have proposed a natural equivalence relation for strong SSPT phases, and correspondingly a classification of a particular class of strong SSPT phases.
Phases with subsystem symmetries differ in key ways from those with traditional global symmetries, necessitating these additional tools.
There are various pertinent directions for future work.

The 2D models with linear (1D) subsystem symmetries studied here are the simplest cases of SSPTs.
Various other SSPTs exist for which it is not obvious how our construction generalizes.
For example, it remains unclear whether there is a meaningful distinction between strong and weak fractal SSPTs~\cite{Devakul2018-ru,Kubica2018-dp} in 2D.
One also has 3D SSPTs with linear subsystem symmetries~\cite{You2018-em}, which are more complicated and for which one must again specify the meaning of a weak SSPT.  

Perhaps of more interest are 3D SSPTs protected by planar (2D) subsystem symmetries, which are mapped to fracton topological orders~\cite{You2018-em} after gauging the symmetries.  
An appealing equivalence relation for strong SSPTs of this type would, on the gauge dual, coincide with the foliated fracton phases~\cite{Shirley2017-fi,Shirley2018-bx,Shirley2018-jy,Shirley2018-yj,Shirley2018-en} (although the connection is not clear at the present).
However, in Ref.~\onlinecite{Shirley2018-yj}, the semionic X-cube was shown to be in the same foliated fracton phase as the regular X-cube, which therefore implies that, on the dual side, the SSPT presented in Ref.~\onlinecite{You2018-em} should be categorized as weak (if the defintions of strong SSPT and foliated fracton phase indeed coincide).
There are also certain fracton models~\cite{Prem2018-nv,Song2018-du} for which it is not clear how they fall into the foliated classification.
Noticing that the strong linear SSPT phase is characterized by non-trivial 2-cocycles between symmetries from different rows, we may conjecture that a strong 3D planar SSPT should be characterized by non-trivial 3-cocycles between symmetries from different planes.
Such a model, if it is indeed strong (and therefore cannot be made up by stacks of 2D SPTs), would result in interesting fracton phases when gauged~\cite{Shirley2018-en}.


\begin{acknowledgements}
TD would like to thank Fiona Burnell and Shivaji Sondhi for support and discussions stimulating this work, as well as Michael Zaletel and Nick Bultinck for various useful discussions, and Wilbur Shirley and Xie Chen for a helpful explanation of their work. 
DW thanks Fiona Burnell, Xie Chen, Meng Cheng, Arpit Dua, and Michael Hermele for interesting discussions. 
TD was supported in part by the National Science Foundation Grant No. NSF PHY-1748958.
YY is supported by PCTS Fellowship at Princeton University.
\end{acknowledgements}

\begin{appendix}
\section{Computing $\mathcal{C}[G]$}\label{app:comp}
Here we show how to compute the classification $\mathcal{C}[G]$ for a general finite abelian group $G$.
By the fundamental theorem of finite abelian groups, a general finite abelian group $G$ may be written as
\begin{equation}
    G = \prod_{i} \mathbb{Z}_{n_i}
\end{equation}
where $n_i$ are prime powers, and $i=1,\dots,N$ for some finite $N$.
The second cohomology group for $G$ is obtained by 
applying the Kunneth formula (for this particular case, see for example the Appendix of Ref.~\onlinecite{Potter2016-zs}),
\begin{equation}
    \mathcal{H}^2[G,U(1)] = \prod_{i<j} \mathbb{Z}_{\gcd(n_i,n_j)}\label{eq:cohom}
\end{equation}
Applying Eq.~\ref{eq:cohom} to the group $G^2$ instead, we get
\begin{equation}
\mathcal{H}^2[G^2,U(1)] = \left(\prod_{i<j}[\mathbb{Z}_{\gcd(n_i,n_j)}]^4\right)\left(\prod_i \mathbb{Z}_{n_i}\right)\label{eq:g2cohom}
\end{equation}

Finally, we wish to compute 
\begin{equation}
\mathcal{C}[G] = \mathcal{H}^2[G^2,U(1)]/(\mathcal{H}^2[G,U(1)])^{3}
\end{equation}
which is easily obtained from Eq.~\ref{eq:cohom} and Eq.~\ref{eq:g2cohom},
\begin{align}
\begin{split}
\mathcal{C}[G] &= \left(\prod_{i<j}\mathbb{Z}_{\gcd(n_i,n_j)}\right)\left(\prod_i \mathbb{Z}_{n_i}\right)\\
 &= \prod_{i\leq j}\mathbb{Z}_{\gcd(n_i,n_j)}
\end{split}
\end{align}

\end{appendix}



 


\bibliography{references}
\end{document}